%% file: main_arxiv.tex
\documentclass[11pt, a4paper, logo, copyright]{googledeepmind}

\usepackage[authoryear, sort&compress, round]{natbib}

\input{math_commands.tex}
\usepackage{times}
\usepackage{latexsym}
\usepackage[T1]{fontenc}
\usepackage[utf8]{inputenc}
\usepackage{microtype}
\usepackage{inconsolata}
\usepackage{graphicx}
\usepackage{balance}       
\usepackage{graphics}      
\usepackage{hyperref}
\usepackage{color}
\usepackage{booktabs}
\usepackage{textcomp}
\usepackage{subcaption}
\usepackage{enumerate}
\usepackage{xcolor}
\usepackage{lipsum}
\usepackage{makecell}
\usepackage{multicol}
\usepackage{multirow}
\usepackage{array}
\usepackage{verbatimbox}
\usepackage{enumitem}
\usepackage{amsmath}
\usepackage{float}
\usepackage{stfloats}
\usepackage{graphicx}
\usepackage{pdflscape}
\usepackage{amsthm}
\usepackage{listings}
\usepackage{caption} 
\usepackage{cleveref}
\usepackage[export]{adjustbox}
\usepackage{xspace}
\usepackage{epsfig}
\usepackage[linesnumbered]{algorithm2e}
\usepackage{algpseudocode}
\usepackage{tabularx}
\usepackage{arydshln}
\usepackage[bottom]{footmisc}
\usepackage{tcolorbox}
\usepackage{stackengine}
\usepackage{placeins}
\usepackage{longtable}
\usepackage{rotating} 
\usepackage[normalem]{ulem}
\useunder{\uline}{\ul}{}
\tcbuselibrary{breakable}

\definecolor{E+F}{RGB}{	255, 99, 71}
\definecolor{B+F}{RGB}{255, 165, 0}
\definecolor{E+I}{RGB}{	173, 216, 230}
\definecolor{B+I}{RGB}{	30, 144, 255}
\definecolor{D}{RGB}{	60, 179, 113}

\definecolor{maroon}{cmyk}{0,0.87,0.68,0.32}

\usepackage{tikz}

\definecolor{darkgreen}{rgb}{0.0, 0.5, 0.0}
\definecolor{usercolor}{RGB}{200, 230, 250} 
\definecolor{coachcolor}{RGB}{180, 250, 180} 

\title{Beyond BMI: Smartphone Body Composition Phenotyping for Cardiometabolic Risk Assessment}
\reportnumber{} 
\renewcommand{\today}{2026-02-01}

\author[1$\dagger$,$\star$]{Menglian Zhou}
\author[1$\dagger$]{Arno Charton}
\author[1]{Emily Blanchard}
\author[1]{Lawrence Cai}
\author[1]{Tracy Giest}
\author[1]{Herschel Watkins}
\author[1]{Mohamed Bouterfa}
\author[1]{Seobin Jung}
\author[1]{Jackie Wasson}
\author[1]{Keerthana Natarajan}
\author[1]{Aniket Deshpande}
\author[1$\S$]{Jiening Zhan}
\author[1$\S$]{Shelten Yuen}
\author[1$\S$]{Javier L. Prieto}
\author[1]{Jacqueline Shreibati}
\author[1]{Mark Malhotra}
\author[1]{Shwetak Patel}
\author[1]{Lindsey Sunden}
\author[1$\S$]{Cathy Speed}
\author[1]{Alicia Kokoszka}
\author[1]{Aravind Natarajan}
\author[1]{Alexandros Pantelopoulos}
\author[1$\star$]{Ahmed A. Metwally}

\affil[1]{Google Research, Mountain View, CA}
\affil[$\dagger$]{Equal Contribution}
\affil[$\S$]{Work done while at Google}
\affil[$\star$]{Correspondence: cassiezhou@google.com and aametwally@google.com}

\correspondingauthor{cassiezhou@google.com, aametwally@google.com}

\input{sections/0-abstract}
\begin{document}

\maketitle
\input{sections/1-main}
\input{sections/2-results}
\input{sections/3-discussion}
\input{sections/4-methods}

\section{Disclaimer}

This paper presents a foundational research study exploring the statistical correlations between novel, accessible body composition estimation methods and established cardiometabolic health markers. The "PhotoScan" and BIA technologies described herein are investigational tools developed for research purposes only and are not commercial products or medical devices.

The models and findings are intended to contribute to the scientific understanding and are not validated for clinical use, do not provide a medical diagnosis, and should not be used for self-diagnosis or to make any health-related decisions. The discussion of predicting risk for conditions like insulin resistance is a purely academic exploration of statistical associations. Any real-world application intended to screen, diagnose, or predict a medical condition would be subject to separate clinical validation and regulatory analysis.

\section{Acknowledgement}

We are deeply grateful to the study participants who contributed their data to this research. We thank Ray Luo, Miao Liu, Jacob Gile, and the entire Human Research Laboratory at Google for their spectacular work in recruiting the study cohorts and collection of multimodality data.

\section{Data Availability}

Data and code used in this study will be made publicly available upon the acceptance of this manuscript

\section{Competing Interests}

This study was funded by Google LLC. MZ, AC, EB, LC, TG, MF, SJ, JW, KN, AD, JS, MM, SP, LS, AK, AN, AP, and AAM are employees of Alphabet and may own stock as part of the standard compensation package.

\pagebreak

\bibliography{main_arxiv}

\pagebreak

\section*{Supplementary Figures}
\input{supplementary/supplementary_figures}

\pagebreak

\section*{Supplementary Tables}
\input{supplementary/supplementary_tables}

\end{document}

%% file: sections/0-abstract.tex
\begin{abstract}

Body Mass Index (BMI) is a widely accessible but imprecise proxy of cardiometabolic health. While assessing true body composition is superior, gold-standard methods like Dual-Energy X-ray Absorptiometry (DXA) are not scalable. We address this gap by developing and validating "PhotoScan," a method to estimate body composition from smartphone imagery. We pretrained  a deep learning model on UK Biobank participants (N=35,323) and fine-tuned on a newly recruited clinical cohort (PhotoBIA cohort, N=677) with diverse ethnicity, age, and body fat distribution, achieving high accuracy against DXA for total body fat percentage (BF{\%}, MAE = 2.15{\%}), Android-to-Gynoid fat ratio (A/G, MAE = 0.11), and visceral-to-subcutaneous fat area ratio (V/S, MAE = 0.09). Generalizability of the model was demonstrated on an independent metabolic health study cohort (MetabolicMosaic cohort, N=132 participants), achieving MAEs of 2.13{\%} for BF{\%}, 0.09 for A/G, and 0.09 for V/S. We then evaluated the clinical utility of these metrics in the MetabolicMosaic cohort by predicting insulin resistance (IR). Adding PhotoScan-derived body composition metrics to baseline demographics model (Age, Sex, BMI) significantly improved insulin resistance classification (Area Under the Receiver Operating Characteristic Curve “AUROC” 76.0{\%} vs 69.2{\%}, DeLong test p=0.002, Net Reclassification Index “NRI” 0.593). Crucially, this accessible smartphone method achieved performance nearly equivalent to adding clinical-grade DXA data to baseline demographics model (AUROC 77.3{\%} vs 69.2{\%}, DeLong test p=0.004, NRI 0.748). These findings demonstrate that smartphone-based phenotyping captures clinically meaningful risk signals missed by BMI and anthropometrics, offering a scalable alternative to DXA for cardiometabolic risk stratification.

\end{abstract}

%% file: sections/1-main.tex
\section{Main}
Cardiometabolic health encompasses two principal components: the cardiovascular system, comprising the heart and blood vessels, and the metabolic processes of the body, which govern energy utilization. An optimal state of cardiometabolic health is typically delineated by favorable levels of adiposity, blood pressure, blood glucose, cholesterol, and triglycerides, alongside the absence of prior pertinent clinical incidents ~\citep{LloydJones2010}. Furthermore, extensive evidence suggests that optimal cardiometabolic health is characterized by a constellation of healthy behaviors, including regular exercise ~\citep{Lin2015}, a balanced diet ~\citep{Wang2023}, abstinence from smoking ~\citep{Leopold2021}, and adequate sleep ~\citep{Knutson2010}. The societal and economic consequences of poor cardiometabolic health are staggering, placing an immense burden on healthcare systems, economies, and families worldwide through the high cost of managing chronic diseases, hospitalizations, and long-term care ~\citep{Ricardo2024}. 

Substantial interest has arisen in technologies and models designed to identify individuals predisposed to cardiometabolic disease, which would enable individuals to adopt straightforward, non-pharmacological wellness strategies, such as dietary modifications and increased physical activity, to mitigate their risk profile ~\citep{Gaesser2015}. While from a clinical point of view the most accurate way for screening for insulin resistance relies on accurate insulin resistance determination ~\citep{Tahapary2022}, there is a growing need for more accessible and less invasive screening tools, leveraging the power of advanced technologies and models utilizing a plethora of longitudinal sensor data collected from a variety of ubiquitous consumer electronic devices ~\citep{Metwally2024, Metwally2026}. 

The composition of the human body is intricately linked to cardiometabolic health, with the distribution and proportion of fat versus lean mass emerging as critical determinants of disease risk. Increases in total body fat percentage (BF{\%}), total visceral fat mass, and the Android fat {\%} divided by the Gynoid fat {\%} (A/G) are all positively and significantly associated with a higher prevalence of insulin resistance ~\citep{Shao2023,Rubino2025,Okosun2015,ZegarraLizana2019,Kurniawan2018}. The impact of these associations surpasses that of traditional metrics like BMI, which is calculated simply from weight and height, fails to differentiate between fat mass and lean mass, nor does it account for the crucial distribution of adipose tissue throughout the body. For instance, an individual with a high BMI might possess a substantial amount of muscle mass rather than excessive adiposity, leading to an inaccurate classification of their metabolic risk. This can result in either underestimating or overestimating an individual's true susceptibility to metabolic disorders.
There is a growing consensus that a more detailed analysis of body composition offers a significantly deeper and more accurate understanding of an individual's nutritional status, disease risk, and physiological changes over time ~\citep{Shao2023,Bennett2025}. Advancements in body composition measurement techniques, including Dual-energy X-ray Absorptiometry (DXA) ~\citep{Laskey1996,Krugh2025}, Bioelectrical Impedance Analysis (BIA) ~\citep{Kushner1992}, and particularly Magnetic Resonance Imaging (MRI) ~\citep{Ross2000}, have enabled large-scale population studies that provide these granular insights, paving the way for more precise identification of at-risk individuals and the development of targeted interventions. The accuracy of new accessible methods is often benchmarked against DXA, whose standard error (SEE) of BF{\%} estimate is 1.6 compared against the 4-Compartment model ~\citep{VanDerPloeg2003} and repeatability (TEM, technical error of measurement) below 1{\%} with strict positioning protocols ~\citep{Nana2015}.
Over the past two decades, a wide range of accessible BIA devices have been introduced to the market, including weight scales, hand-held units, and wearable technologies such as watches.  In recent years, several wrist-worn wearable devices have been introduced to the market, including the Samsung Galaxy watches ~\citep{Bennett2022,Jung2021}, the Asus Vivowatch, and the AURA Strap, which is a wrist-wearable Apple Watch accessory ~\citep{Polokhin2022}. Another distinct technological method for estimating detailed body composition, particularly fat distribution, involves capturing human body images (with minimal clothing to ensure visibility of muscle definition and detailed fat distribution) and subsequently training deep learning models to extract complex embeddings from the human silhouette. These embeddings are then mapped to metrics such as android and gynoid fat. This methodology demonstrated an MAE of 2.16{\%} for BF{\%} in a study comprising 134 participants ~\citep{Majmudar2022}. Similarly, ~\citep{Choudhary2023} showed that smartphones combined with advanced computer vision algorithms can provide relatively accurate estimation of waist circumference.

While numerous studies have established strong associations between waist circumference, body fat percentage, and their respective associations with insulin resistance, there is a critical gap in comprehensively investigating these anthropometric and body composition measures, especially when derived from emerging technologies like 3D photogrammetry, concurrently to quantify their relative importance in predicting insulin resistance. In this paper, we (1) recruited and curated diverse multimodal cohorts (DXA, demographics, anthropometrics, BIA, and PhotoScan), enabling comprehensive model pretraining, fine-tuning, and independent validation of body composition (BF{\%}, A/G, and V/S) and insulin resistance prediction, (2) introduced and validated a method for predicting body composition directly from participant-derived 2D body scans (PhotoScan, images captured by smartphone from two views, two camera heights and two poses), where our model achieves state-of-the-art predictive accuracy, (3) identified the relative importance of incorporating waist circumference and body composition (derived from DXA, PhotoScan, and BIA) in enhancing the prediction of insulin resistance, (4) publicly releasing the complete dataset used in this study and open-sourcing all code used in model training and evaluation, which will have tremendous value to the scientific community working on cardiometabolic health and beyond.

%% file: sections/2-results.tex
\section{Results}
\label{sec:results}

\subsection{Study design and cohort characteristics}
This study included three cohorts. The first cohort is the MetabolicMosaic Cohort, where we enrolled 153 participants, with an emphasis on individuals who were overweight or obese, into a 30-week prospective longitudinal study. This cohort was used to train and test models for predicting insulin resistance using body composition (from ground-truth DXA, smartphone-based PhotoScan, or smartwatch-based BIA), wearable data, anthropometrics (e.g., waist circumference), and demographics (age, sex, ethnicity). The study was approved by the WCG Institutional Review Board (IRB \#1371945; Methods). At the beginning and end of the study, participants attended in-person appointments at the Google Human Research Laboratory in San Francisco for anthropometric measurements, blood pressure measurements, blood tests, and body composition measurements (Methods). Additionally, participants were instructed to wear a Fitbit Charge 6 for a minimum of 20 hours per day throughout the study for the passive collection of activity, sleep, and heart rate-derived metrics. At the study's conclusion, we obtained 195 complete data sessions (from first and/or last visits) from 132 unique subjects (\Cref{fig:data_processing_workflow}). The cohort distribution consisted of 54 insulin-sensitive, 46 insulin-impaired, and 32 insulin resistant individuals. The detailed subject-level cohort characteristics are shown in \Cref{fig:metabolicmosaic_cohort} and \Cref{tab:metabolic_mosaic}, a detailed breakdown by visit is available in \Cref{tab:cohort_characteristics}.

\begin{table}[tbp]
\centering
\captionsetup{justification=centering}
\caption{MetabolicMosaic Study Cohort Characteristics}
\label{tab:metabolic_mosaic}
\begin{tabular}{lccccc}
\toprule
& \textbf{ALL} & \multicolumn{3}{c}{\textbf{Insulin Resistance}} & \textbf{P-Value} \\
\cmidrule(lr){3-5}
\textbf{Characteristics} &  & \textbf{IS} & \textbf{IMPAIRED-IS} & \textbf{IR} & \\
\midrule
\textbf{Sample size} & & & & & \\
\# of Participants & 132 & 54 & 46 & 32 & - \\
\# of Sessions & 195 & 84 & 64 & 47 & - \\
\midrule
\multicolumn{6}{l}{\textbf{Gender}} \\
Female & 131 & 54 & 46 & 31 & 0.61 \\
Male & 64 & 30 & 18 & 16 & 0.61 \\
\midrule
\multicolumn{6}{l}{\textbf{Ethnicity}} \\
Asian & 82 & 36 & 28 & 18 & 0.008 \\
Black or African American & 11 & 2 & 3 & 6 & 0.008 \\
Hispanic or Latino & 16 & 2 & 5 & 9 & 0.008 \\
White & 66 & 33 & 21 & 12 & 0.008 \\
More than one race & 18 & 11 & 6 & 1 & 0.008 \\
Other & 2 & 0 & 1 & 1 & 0.008 \\
\midrule
\multicolumn{6}{l}{\textbf{Demographics}} \\
Age & 43.4 (11.4) & 43.1 (11.3) & 43.2 (10.8) & 44.3 (12.6) & 0.83 \\
BMI & 29.0 (7.1) & 26.0 (5.4) & 29.1 (6.4) & 34.1 (7.6) & $< 0.001$ \\
\midrule
\multicolumn{6}{l}{\textbf{Blood Biomarkers}} \\
HDL & 51.6 (15.7) & 57.8 (15.4) & 50.0 (13.3) & 42.6 (14.6) & $< 0.001$ \\
HgA1c & 5.1 (0.7) & 5.1 (0.7) & 5.1 (0.5) & 5.3 (0.8) & 0.2 \\
HOMA-IR & 2.1 (1.5) & 1.0 (0.3) & 2.1 (0.4) & 4.3 (1.5) & $< 0.001$ \\
LDL & 124.2 (33.4) & 126.7 (29.7) & 124.3 (36.2) & 120.2 (35.3) & 0.6 \\
TAG & 107.9 (59.9) & 90.1 (42.1) & 106.3 (52.5) & 137.5 (79.0) & $< 0.001$ \\
FSBG & 93.0 (10.4) & 88.8 (7.8) & 92.9 (8.4) & 100.8 (12.5) & $< 0.001$ \\
Total Cholesterol & 193.4 (40.0) & 197.7 (32.7) & 188.6 (46.8) & 192.4 (42.0) & 0.38 \\
\midrule
\multicolumn{6}{l}{\textbf{Body Comp}} \\
Waist & 90.7 (15.6) & 83.7 (12.6) & 90.8 (14.5) & 103.2 (14.4) & $< 0.001$ \\
Hip Circumference & 106.3 (14.7) & 101.2 (11.2) & 105.5 (13.9) & 116.5 (16.3) & $< 0.001$ \\
Body Fat (BF) \% & 34.4 (8.2) & 31.9 (7.6) & 34.7 (7.8) & 38.4 (8.3) & $< 0.001$ \\
A/G* & 0.5 (0.2) & 0.4 (0.1) & 0.5 (0.1) & 0.6 (0.2) & $< 0.001$ \\
V/S** & 0.3 (0.1) & 0.3 (0.1) & 0.3 (0.1) & 0.3 (0.1) & 0.58 \\
Blood Pressure (Diastolic) & 76.5 (8.2) & 73.5 (7.3) & 77.4 (7.6) & 80.7 (8.6) & $< 0.001$ \\
\bottomrule
\multicolumn{6}{l}{\footnotesize *A/G: Android Fat / Gynoid Fat ; ** V/S: Visceral Fat Area / Abdomen Subcutaneous Fat Area} \\
\end{tabular}%
\end{table}

The second cohort is the UK Biobank (UKBB) Cohort, which was used to pre-train PhotoScan models for body composition prediction. This study was conducted using the UK Biobank Resource under Application Number 65275 (Methods). The subset of the UKBB containing both MRI images and DXA-derived BF\% ground truth included 35,323 participants (18,273 female, 17,050 male; median age: 63.8 years; median BMI: 26.6 kg/m\textsuperscript{2}) (\Cref{tab:ukbb_characteristics}).

\begin{table}[tbp]
\centering
\captionsetup{justification=centering}
\caption{UK Biobank Dataset Cohort Characteristics}
\label{tab:ukbb_characteristics}
\begin{tabular}{lccc}
\toprule
\textbf{UK Biobank} & \textbf{All} & \textbf{Male} & \textbf{Female} \\
\midrule
N & 35323 & 17050 & 18273 \\
Age & 63.8 (7.6) & 64.5 (7.7) & 63.2 (7.4) \\
BMI & 26.6 (4.3) & 27.1 (3.9) & 26.1 (4.6) \\
Waist circumference & 88.4 (12.6) & 94.5 (10.5) & 82.8 (11.7) \\
Body Fat (BF) \% & 35.0 (8.0) & 30.4 (6.3) & 39.3 (7.0) \\
\bottomrule
\end{tabular}
\end{table}

The third cohort we used is the PhotoBIA Cohort, which is a newly recruited study (N=677) and was used for two purposes: (1) fine-tuning the PhotoScan body composition models and (2) training and testing the BIA models for BF\% prediction. This study was approved under IRB protocol Pro00065782 (Methods) and was conducted to collect development data for these estimation models. The cohort included 677 participants (385 female, 289 male, 3 other gender identity; median age: 41.9 years; median BMI: 26.2 kg/m\textsuperscript{2}) (\Cref{tab:photobia_characteristics}). To minimize potential impacts on BIA and DXA measurements, participants were instructed not to eat or drink for four hours prior to their lab appointments (Methods).

\begin{table}[tbp]
\centering
\captionsetup{justification=centering}
\caption{PhotoBIA Cohort Characteristics}
\label{tab:photobia_characteristics}
\begin{tabular}{lccc}
\toprule
\textbf{Characteristics} & \textbf{All} & \textbf{Male} & \textbf{Female} \\
\midrule
\textbf{N} & 677 & 289 & 385 \\
\midrule
\multicolumn{4}{l}{\textbf{Demographics}} \\
Age (years) & 41.9 (13.9) & 43.2 (14.7) & 41.2 (13.3) \\
BMI ($kg/m^2$) & 26.2 (6.6) & 27.1 (5.8) & 25.4 (7.0) \\
\midrule
\multicolumn{4}{l}{\textbf{Ethnicity (N)}} \\
White & 244 & 116 & 127 \\
Asian & 167 & 71 & 96 \\
African American & 91 & 37 & 52 \\
Hispanic & 81 & 35 & 45 \\
Other & 96 & 30 & 64 \\
\midrule
\multicolumn{4}{l}{\textbf{Body Composition}} \\
Waist circumference (cm) & 84.3 (15.0) & 90.9 (13.4) & 79.2 (14.2) \\
Abdomen circumference (cm) & 90.3 (16.1) & 94.5 (15.1) & 87.1 (16.0) \\
Body Fat (BF) \% & 29.9 (8.2) & 24.9 (6.1) & 33.7 (7.4) \\
\bottomrule
\end{tabular}%
\end{table}

\begin{figure}[tbp]
    \centering
    \includegraphics[width=\linewidth]{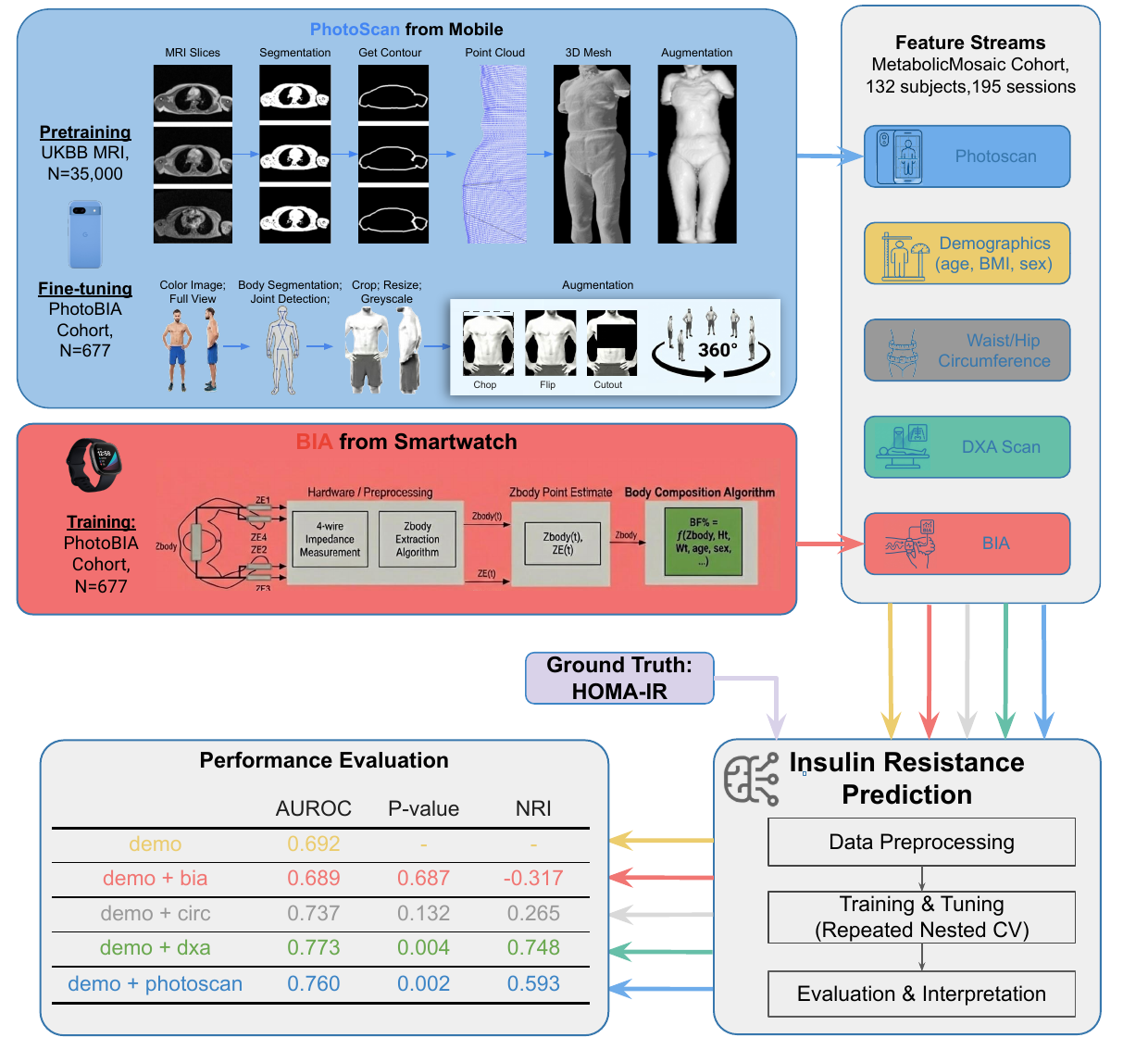}
    \caption{\textbf{Overview of the Body Composition Training Pipeline and Insulin Resistance Model Performance.} The PhotoScan model was pre-trained using the UK Biobank cohort and fine-tuned on the PhotoBIA cohort, while the BIA model was trained on the PhotoBIA cohort. Body composition inferences on the MetabolicMosaic cohort (derived from both PhotoScan and BIA models), combined with DXA estimates and anthropometric measurements, were consolidated as feature sets to train independent insulin resistance classifiers. Model performance metrics are summarized in the inset (bottom left). The DeLong test p-values indicate whether the model's AUC differs significantly from the demographics-only baseline. The Net Reclassification Index (NRI) quantifies the improvement in correctly classifying subjects relative to the baseline.}
    \label{fig:pipeline}
\end{figure}

\begin{figure}[tbp]
    \centering
    \includegraphics[width=\linewidth]{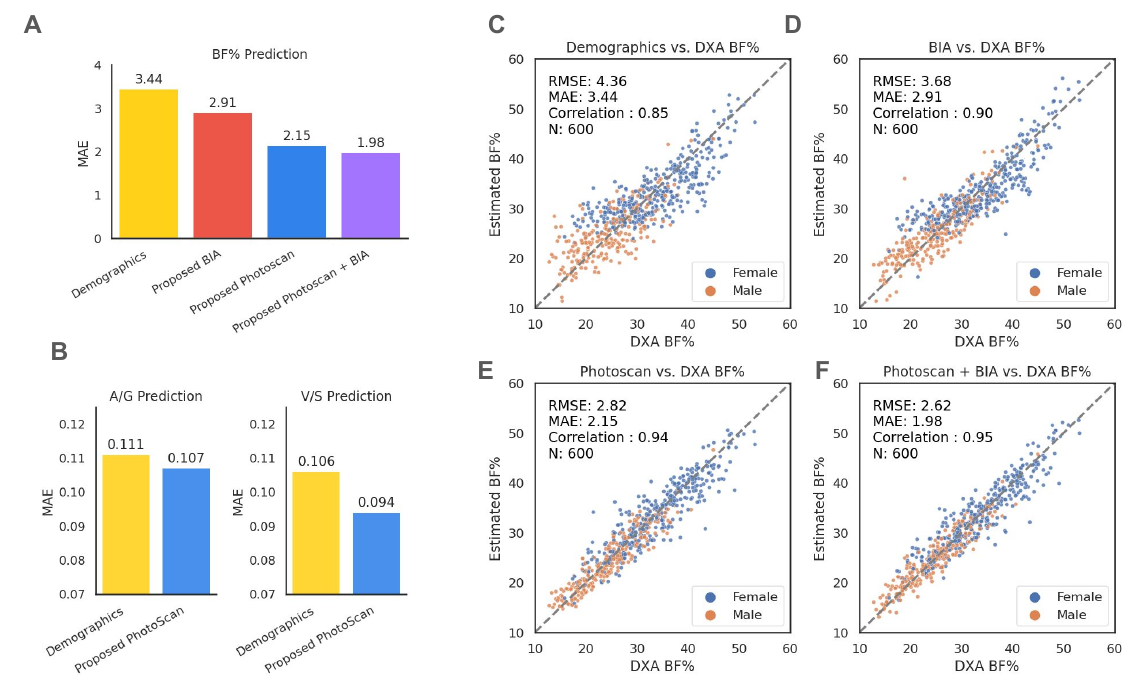}
    \caption{\textbf{Body Composition Model Performance on the PhotoBIA Cohort.} A) Comparative accuracy of inferred BF\% against ground truth (DXA-derived BF\%) for demographics-only model, BIA model, PhotoScan model and BIA and PhotoScan fusion model. B) Comparative accuracy of inferred android to gynoid fat percentage ratio (A/G) and visceral to subcutaneous fat area ratio (V/S) against ground truth (DXA-derived A/G) for demographics-only model and PhotoScan model. C - F) Scatterplot of BF\% predictions from C) demographics only model D) BIA model E) PhotoScan model F) BIA-PhotoScan fusion model, against ground truth (DXA-derived BF\%)}
    \label{fig:photobia_perf}
\end{figure}

\begin{figure}[tbp]
    \centering
    \includegraphics[width=\linewidth]{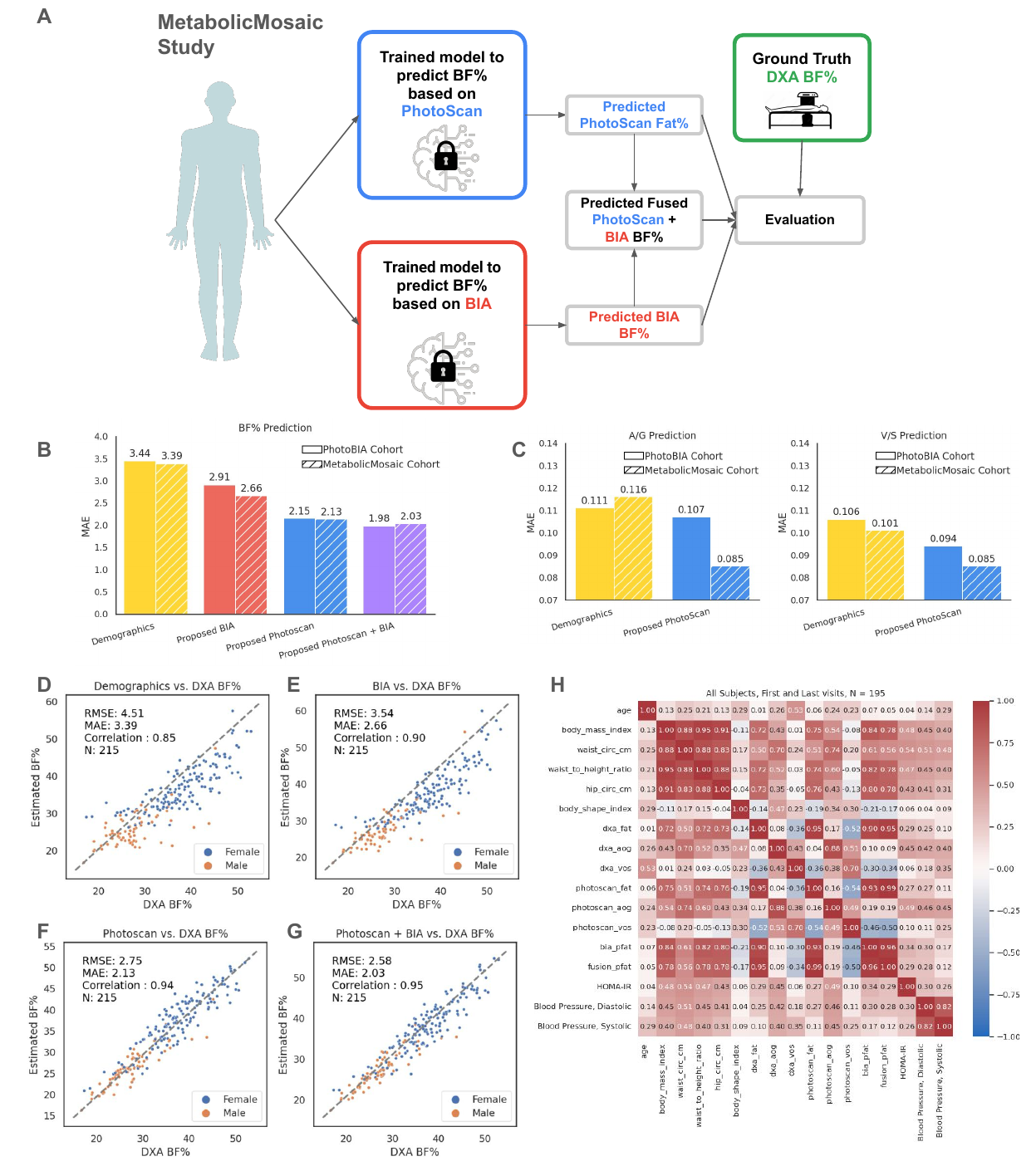}
    \caption{\textbf{Inference of Body Composition on the MetabolicMosaic Cohort, and the Association with Cardiometabolic Health Metrics.} A) Overview of the inference and evaluation pipeline for BF\%. Overall, there are 215 visits with both valid PhotoScan and BIA measurements. B) Comparative accuracy of inferred BF\% against ground truth (DXA-derived BF\%) for demographics-only model, BIA model, PhotoScan model and BIA and PhotoScan fusion model for PhotoBIA Cohort and MetabolicMosaic cohort. C) Comparative accuracy of inferred A/G and V/S against ground truth (DXA-derived A/G and V/S) for demographics-only model and PhotoScan model for PhotoBIA Cohort and MetabolicMosaic cohort. D - G) Scatter Plot of ground truth (DXA-derived BF\%) against BF\% predictions from D) demographics-only model E) BIA model F) PhotoScan model G) BIA-PhotoScan fusion model.}
    \label{fig:mosaic_inference}
\end{figure}

The study design, schematically represented in \Cref{fig:pipeline}, employed a comprehensive approach to predict insulin resistance using an array of multimodal data sources (see Methods). We trained a suite of machine learning models on this integrated dataset. The dataset included five distinct data modalities including demographic and anthropometric features such as age, BMI, sex, and precise waist and hip circumference measurements and their ratios; gold-standard body composition data derived from DXA, encompassing metrics like BF\%, visceral fat percentage, A/G and V/S; and mobile-based body composition estimates obtained via the PhotoScan technology, which yielded BF\%, fat distribution, and BIA values, as well as a fused measure combining PhotoScan and BIA-predicted BF\%. Finally, a panel of crucial blood biomarkers, such as a full lipid panel, glycated hemoglobin (HbA1c), and fasting glucose, are recorded to provide a comprehensive characterization of the cohort. The specific features comprising each modality are fully detailed in \Cref{tab:feature_sets}. The development and extensive validation of the trained models for predicting body composition from PhotoScan and BIA are described in the following sections of this work (\Cref{fig:photobia_perf} and \Cref{fig:mosaic_inference}). Our classification strategy involved a systematic assessment of each individual data modality and their combinations to identify the most potent feature set for predicting IR.

\begin{table}[tbp]
\centering
\captionsetup{justification=centering}
\caption{Feature Sets and Variable Descriptions}
\label{tab:feature_sets}
\resizebox{\textwidth}{!}{%
\begin{tabular}{llll}
\toprule
\textbf{Category} & \textbf{Feature Set} & \textbf{Feature} & \textbf{Feature Short Name} \\
\midrule
\multirow{8}{*}{\begin{tabular}[c]{@{}l@{}}Anthropometric\\ Measurements\end{tabular}} & \multirow{3}{*}{demo} & Chronological age & age \\
 & & Race/ethnicity & race \\
 & & Body Mass Index (BMI) & body\_mass\_index \\
\cmidrule(lr){2-4}
 & \multirow{5}{*}{circ} & Waist circumference (cm) & waist\_circ\_cm \\
 & & Hip circumference (cm) & hip\_circ\_cm \\
 & & Waist circumference to hip circumference ratio & waist\_to\_hip\_ratio \\
 & & Waist circumference to height ratio & waist\_to\_height\_ratio \\
 & & \begin{tabular}[c]{@{}l@{}}Waist circumference divided by the product of BMI to the\\ power of two-thirds and the square root of height\end{tabular} & body\_shape\_index \\
\midrule
\multirow{7}{*}{\begin{tabular}[c]{@{}l@{}}Body Composition\\ Measurements\end{tabular}} & \multirow{3}{*}{dxa} & Body fat percentage (from DXA) & dxa\_fat \\
 & & Android-to-gynoid fat ratio (from DXA) & dxa\_aog \\
 & & Visceral-to-subcutaneous fat area ratio (from DXA) & dxa\_vos \\
\cmidrule(lr){2-4}
 & \multirow{3}{*}{photoscan} & Body fat percentage (from PhotoScan) & photoscan\_fat \\
 & & Android-to-gynoid fat ratio (from PhotoScan) & photoscan\_aog \\
 & & Visceral-to-subcutaneous fat area ratio (from PhotoScan) & photoscan\_vos \\
\cmidrule(lr){2-4}
 & bia & Body fat percentage (from BIA) & bia\_fat \\
\bottomrule
\end{tabular}%
}
\end{table}

For insulin resistance classification, we used HOMA-IR ([Fasting Insulin ($\mu$U/ml) $\times$ Fasting Glucose (mg/dL)]/ 405), as the ground truth. As defined in [10] participants were classified as insulin resistant (IR) (HOMA-IR $>$ 2.9), insulin sensitive (IS) (HOMA-IR $<$ 1.5), or impaired insulin sensitivity (Impaired-IS) ($1.5 \le \text{HOMA-IR} \le 2.9$). From first and last visits, in total we have 47 sets of data with insulin resistance and 148 sets without insulin resistance (i.e., insulin sensitive and impaired insulin sensitivity) (\Cref{tab:metabolic_mosaic}).

We trained models using repeated nested cross validation framework, and we performed a comprehensive evaluation of the predicted insulin resistance classes to assess the generalizability. Moreover, we performed interpretability analysis on the trained models to identify feature importance. Specifically, the inclusion of circumference measurements combined with body composition data (derived from either mobile PhotoScan or DXA) yielded a substantial and statistically significant improvement in overall model accuracy compared to demographics (\Cref{fig:pipeline}). This enhancement demonstrates the critical role of physical morphology in capturing variability beyond simple demographic information.

\subsection{Predicting Body composition via PhotoScan from Smartphone Camera}
The PhotoScan model is a deep learning framework that infers three body composition metrics, BF\%, A/G, and visceral to subcutaneous fat area ratio (V/S) from a pair of frontal and lateral smartphone images (\Cref{fig:pipeline}). The model was developed using a two-stage process: pretraining on UK Biobank MRI data with DXA-derived ground truth (N=35,323; Methods, \Cref{tab:ukbb_characteristics}), followed by fine-tuning on our PhotoBIA cohort using smartphone images (N=677; Methods, \Cref{tab:photobia_characteristics}).

We first pretrained a model to predict the three body composition metrics using MRI 3D data from the UKBB cohort, with corresponding DXA scans providing the ground truth values. After extensive preprocessing of the MRI data (Methods), 2D projections representing frontal and lateral views were generated and combined into single 300x300 pixel images. A standard ResNet-50 deep neural network, initialized with ImageNet weights, was trained on these image projections. The network's inputs included the generated images along with participant sex, height, and weight; BMI was computed internally. The features extracted by the ResNet-50 backbone were concatenated with sex and BMI and processed through a final dense layer with 64 units. A softmax activation function was applied to produce a probability density function for the target metric. The dataset was partitioned into an 80\% training set and a 20\% test set. This pretrained model using MRI images achieved a Mean Absolute Error (MAE) of 2.01\% for BF\% prediction on the UKBB test set.

The pretrained model was subsequently fine-tuned using data from the PhotoBIA cohort, which provided smartphone photos with corresponding DXA-derived ground truth metrics. To augment the training data, we extracted optimal frontal and lateral frames from 360-degree participant videos using a landmark detection pipeline to identify the correct poses automatically. Fine-tuning was conducted using a 5-fold cross-validation strategy, with folds stratified to ensure a balanced distribution of fat percentage values. The resulting fine-tuned models demonstrated an average MAE of 2.15\% (Standard Deviation “SD” = 0.1) across all folds (\Cref{fig:photobia_perf}A). An analysis of the prediction errors for BF\% revealed a near-Gaussian distribution, with a mean of 0.09\% (95\% Confidence Interval: 0.01 – 0.17) and a standard deviation of 2.90\%.

\subsection{Predicting Body Fat Percentage via BIA sensor on Smartwatch}
We also developed a model to predict BF\% using bioelectrical impedance analysis (BIA) measured by an on-wrist smartwatch. For this, we used a research prototype based on the Fitbit Sense 2, which is equipped with a BIA sensor (\Cref{fig:pipeline}) . The sensor employs a 4-wire sensing mechanism at a 50 kHz excitation frequency, utilizing two bottom electrodes, two top electrodes, and a TI AFE4500 chip. Extracting the desired body impedance (Zbody) requires analyzing a high-dimensionality impedance network with 13 unknown parameters: eight parasitic impedances, four contact impedances (Z), and Zbody itself. \Cref{fig:pipeline} illustrates the complete estimation pipeline: a) signal acquisition via the hardware sensor and solving for Zbody, b) deriving a point estimate of Zbody from the BIA time-series, and c) mapping the BIA and demographic data to body composition values (see Methods).

BIA-based model to predict BF\% we trained and tested using linear regression with Tikhonov regularization via 5-fold cross-validation on the PhotoBIA cohort (Methods). The model inputs include the extracted body impedance (as height squared divided by body resistance) along with demographics (age, sex, height, and weight).

\subsection{PhotoScan outperforms BIA and demographic-based models in BF\% prediction}
The resulting fine-tuned PhotoScan model demonstrated an average MAE of 2.15 for BF\% prediction across all folds of the PhotoBIA cohort, while the BIA-based model achieved an MAE of 2.91. Both the PhotoScan and BIA models outperformed a model based solely on demographics (MAE = 3.44), which uses age, sex, height, weight. Fusing (averaging) the predicted BF\% from PhotoScan and BIA achieved a lower MAE than either model alone (MAE = 1.98) (\Cref{fig:photobia_perf}A). Furthermore, models based on PhotoScan showed better performance in predicting specific body fat distributions, A/G and V/S, than demographics alone (MAE = 0.107 vs. 0.111 for A/G, and MAE = 0.094 vs. 0.106 for V/S, respectively) as shown in \Cref{fig:photobia_perf}B. More detailed estimation accuracy is detailed in \Cref{tab:stratified_bf}.

Given that PhotoScan and BIA capture BF\% information via different pathways (visual contour analysis versus electrical impedance), we tested whether a fusion of the two technologies could improve prediction accuracy. Towards that goal, we utilized the BF\% output of the PhotoScan model as an additional feature for the model mapping body impedance and demographics to BF\%. This fused BIA\_PhotoScan model was trained and tested on the PhotoBIA cohort using 5-fold cross-validation, consistent with the BIA-alone model and PhotoScan-alone model. This approach (\Cref{fig:photobia_perf}A) improved BF\% prediction accuracy beyond that of PhotoScan alone, reducing the MAE to 1.98 compared to 2.15.

Moreover, \Cref{fig:photobia_perf}C-F highlight the correlation between ground-truth BF\% from DXA and the predicted BF\% from demographics, BIA, PhotoScan, and the fused PhotoScan-BIA model. These figures also illustrate the effect of sex on the prediction error for each modality. PhotoScan demonstrates a significantly higher correlation with ground truth than demographics alone (Pearson's $r = 0.94$ vs. $0.85$). Notably, the prediction error for males was considerably lower using PhotoScan compared to the demographics-only model.

\subsection{Validation of body composition prediction using PhotoScan and BIA on an independent cohort}
To validate the generalizability of the body composition prediction models developed using PhotoScan and BIA, we evaluated their performance on the MetabolicMosaic Cohort. This cohort is considered an independent validation set, as none of its participants were part of the UKBB or PhotoBIA cohorts used for training (\Cref{fig:mosaic_inference}A). The BF\% estimation accuracy stratified by gender, camera position, pose and ethnicity is shown in \Cref{tab:stratified_bf}, and the A/G, V/S estimation accuracy stratified by gender is shown in \Cref{tab:ag_vs_performance}.

\Cref{fig:mosaic_inference}B shows that the MAE obtained from the frozen models on the independent cohort yielded similar performance to the initial cohorts used for training and testing: BIA (MAE 2.66 vs. 2.91), PhotoScan (MAE 2.13 vs. 2.15), and the fused PhotoScan\_BIA model (MAE 2.03 vs. 1.98). Similarly, for body distribution metrics (\Cref{fig:mosaic_inference}C), the models showed consistent, and slightly better performance for A/G (MAE 0.085 vs. 0.107) and V/S (MAE 0.085 vs. 0.094). 

Similar to the initial cohorts, scatter plots in \Cref{fig:mosaic_inference}D-G show the concordance between the ground-truth DXA BF\% and the predicted BF\% from BIA, PhotoScan, and the fused PhotoScan-BIA model. Although the MAE values were not identical between the initial and independent validation cohorts, the correlation coefficient between the ground-truth and predicted BF\% was exactly the same, demonstrating the model's high generalizability.

\subsection{Body composition metrics from DXA and PhotoScan, and waist circumference, are strongly associated with insulin resistance}
\Cref{fig:mosaic_inference}H shows the pairwise Pearson correlation coefficients between insulin resistance (measured via HOMA-IR), blood pressures (SBP and DBP), body composition from ground-truth DXA (dxa\_fat, dxa\_aog, dxa\_vos) and predicted from PhotoScan (photoscan\_fat, photoscan\_aog, photoscan\_vos), BIA (bia\_fat), fused BF\% from PhotoScan and BIA (fusion\_fat), circumference-based measures (waist\_circ, hip\_circ, waist\_to\_height\_ratio, body shape index), BMI, and age. All correlations, also stratified by sex, and their p-values are listed in \Cref{tab:feature_correlation_all_subjects_final,tab:feature_correlation_female_final,tab:feature_correlation_male_final}.

Significant positive correlations were observed between HOMA-IR and waist circumference ($r=0.54$, $p<0.0001$), BMI ($r=0.48$, $p<0.0001$), waist-to-height ratio ($r=0.47$, $p<0.0001$), hip circumference ($r=0.43$, $p<0.0001$), and the A/G from DXA (dxa\_A/G, $r=0.45$, $p<0.0001$) and from PhotoScan (PhotoScan\_A/G, $r=0.49$, $p<0.0001$). Age, body shape index, BF\% from DXA/PhotoScan/BIA, and the V/S from DXA (dxa\_V/S) and PhotoScan showed weaker correlations with HOMA-IR ($|r| < 0.4$). PhotoScan-based estimates of body composition and distribution (e.g., BF\%, A/G) showed a similar correlation with HOMA-IR compared to ground-truth DXA (\Cref{fig:mosaic_inference}H), indicating that PhotoScan is a promising substitute for DXA in improving insulin resistance prediction.

Recognizing that men and women exhibit different fat storage patterns, with men tending toward an android (``apple'') shape rich in visceral fat and women toward a gynoid (``pear'') shape rich in subcutaneous fat—we also stratified the correlation analysis by sex. The predictive strength of BMI varied significantly, showing a strong correlation with HOMA-IR in females ($r=0.58$, $p<0.0001$) but a weaker, still significant, correlation in males ($r=0.35$, $p<0.0001$). In contrast, the A/G ratio, a direct measure of body shape, was a more consistent metric, maintaining a moderate-to-strong correlation in both females ($r=0.46$, $p<0.0001$) and males ($r=0.55$, $p<0.0001$). These findings indicate that the A/G ratio is a more robust and reliable predictor of insulin resistance than BMI when assessing a mixed-gender population (\Cref{fig:feature_correlation_by_gender}).

For blood pressure, since SBP and DBP were highly correlated ($r=0.82$, $p<0.0001$), we focused this feature correlation analysis on SBP only. We observed significant positive correlations between SBP and waist circumference ($r=0.48$, $p<0.0001$), BMI ($r=0.40$, $p<0.0001$), waist-to-height ratio ($r=0.40$, $p<0.0001$), A/G from DXA ($r=0.40$, $p<0.0001$), and the A/G from PhotoScan ($r=0.45$, $p<0.0001$). Similar to the HOMA-IR findings, correlations between SBP and metrics like BMI, waist circumference, hip circumference, waist-to-height ratio, and BF\% were much lower in male subjects than in female subjects. In contrast, the waist-to-hip ratio had a much higher correlation with SBP in males than in females (\Cref{fig:feature_correlation_by_gender}), a finding likely due to the different fat storage patterns between sexes. Compared to its weak correlation with HOMA-IR ($r=0.04$, $p=0.57$), age had a higher correlation with SBP ($r=0.29$, $p<0.0001$), as SBP tends to increase with age due to progressive arterial stiffening.

\subsection{Prediction of insulin resistance using demographics, body composition and anthropometrics}
We trained models to classify insulin resistance subjects using individual and combined feature sets comprising body composition, anthropometrics, and demographic data. The features included in each set are listed in \Cref{tab:feature_sets}.

For insulin resistance classification, HOMA-IR was the ground truth; subjects with HOMA-IR $>$ 2.9 were categorized as insulin resistance ``IR,'' and those with HOMA-IR $\le 2.9$ were labeled as ``Non-IR'' (defined as insulin sensitive or impaired insulin sensitive). Population characteristics for subjects with and without these conditions are shown in \Cref{tab:metabolic_mosaic} and \Cref{tab:cohort_characteristics}.

\begin{figure}[tbp]
    \centering
    \includegraphics[width=\linewidth]{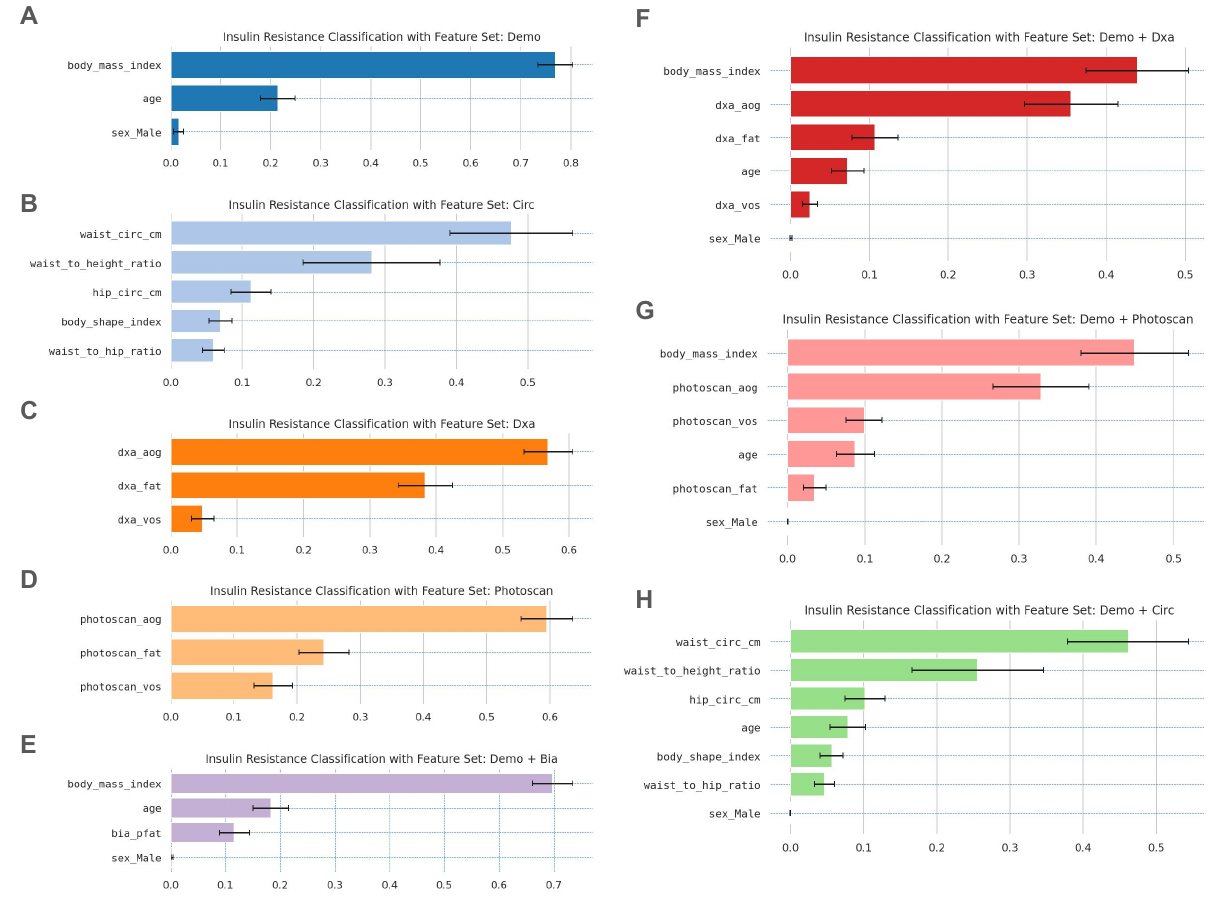}
    \caption{\textbf{Feature Importance for Insulin Resistance Classification Models with Individual Feature Set and Combinations.} Full explanations of each feature set are described in \Cref{tab:feature_importance} A) Demographics Feature Set B) Anthropometric Feature Set C) DXA-derived Body Composition Feature Set. D) PhotoScan-predicted Body Composition Feature Set. E) Demographics combined with BIA Feature Set. F) Demographics combined with DXA-derived Body Composition Feature Set. G) Demographics combined with PhotoScan-predicted Body Composition Feature Set. H) Demographics combined with Anthropometric Feature Sets.}
    \label{fig:feature_importance}
\end{figure}

The insulin resistance classification model was trained using repeated nested cross-validation with an XGBoost classifier with individual and combined feature sets. The cross-validation for insulin resistance was stratified by BMI and insulin resistance labels to ensure each fold was balanced for these factors. Samples from the same subject were placed exclusively in either the training or validation set of any given fold to prevent data leakage. Models were evaluated using both threshold-independent metrics (AUROC and Area Under the Precision-Recall Curve ``AUPRC'') and threshold-dependent metrics (Sensitivity, Precision, F1 score, and Negative Predictive Value ``NPV'') calculated at a fixed specificity of 80\%. Additionally, Net Reclassification Improvement (NRI) and DeLong tests were performed to compare all models against the baseline demographics-only model. As shown in \Cref{tab:model_performance}, incorporating DXA-derived body composition features with demographics yielded the highest AUROC ($0.773$, $p=0.004$) and highest NRI ($0.748$), with AUPRC increased from 0.351 to 0.451 and F1 score at 80\% specificity increased from 0.394 to 0.53. Combining PhotoScan-based body composition metrics with demographics also achieved a high AUROC ($0.760$, $p=0.002$) and NRI ($0.593$), with AUPRC increased to 0.502 and F1 score at 80\% specificity increased to 0.549. Although models using circumference and demographic features did not perform as well as those with body composition features, they still had higher predictive value than models using demographics alone (AUROC $0.737$, $p = 0.132$, AUPRC 0.383 and F1 score 0.453 at 80\% specificity). \Cref{fig:feature_importance} and \Cref{tab:feature_importance} displays the feature importance. Within the demographics feature set, BMI was the dominant feature. Within anthropometrics, waist circumference and waist-to-height ratio were the most important features. For both DXA-derived and PhotoScan-predicted body composition, A/G had the highest feature importance, followed by BF\%. Combining demographics with body composition, BMI had the highest importance, followed by the DXA-derived or PhotoScan-predicted A/G.

\begin{table}[tbp]
\centering
\captionsetup{justification=centering}
\caption{Comparative Analysis of Model Performance Across Varying Feature Sets}
\label{tab:model_performance}
\resizebox{\textwidth}{!}{%
\begin{tabular}{lccccccccc}
\toprule
\textbf{Feature Set} & \textbf{NRI} & \textbf{AUROC} & \textbf{AUPRC} & \textbf{Sens.} & \textbf{Prec.} & \textbf{F1} & \textbf{Spec.} & \textbf{NPV} & \textbf{\makecell{p\_value \\ (DeLong)}} \\
\midrule
demo & 0.692 & 0.351 & 0.400 & 0.388 & 0.394 & 0.800 & 0.808 & - & - \\
circ & -0.023 & 0.738 & 0.394 & 0.502 & 0.445 & 0.472 & 0.801 & 0.835 & 0.160 \\
dxa & -0.031 & 0.748 & 0.436 & 0.587 & 0.479 & 0.528 & 0.797 & 0.859 & 0.195 \\
photoscan & 0.075 & 0.755 & 0.488 & 0.511 & 0.449 & 0.478 & 0.801 & 0.838 & 0.073 \\
bia & -0.581 & 0.605 & 0.325 & 0.426 & 0.403 & 0.414 & 0.800 & 0.814 & \textbf{0.039$^{*}$} \\
demo + circ & 0.265 & 0.737 & 0.383 & 0.477 & 0.431 & 0.453 & 0.800 & 0.828 & 0.132 \\
demo + dxa & 0.748 & 0.773 & 0.451 & 0.587 & 0.483 & 0.530 & 0.800 & 0.859 & \textbf{0.004$^{*}$} \\
demo + photoscan & 0.593 & 0.760 & 0.502 & 0.617 & 0.495 & 0.549 & 0.800 & 0.868 & \textbf{0.002$^{*}$} \\
circ + photoscan & 0.315 & 0.765 & 0.470 & 0.600 & 0.486 & 0.537 & 0.799 & 0.863 & \textbf{0.022$^{*}$} \\
demo + circ + photoscan & 0.352 & 0.764 & 0.448 & 0.570 & 0.475 & 0.518 & 0.800 & 0.854 & \textbf{0.021$^{*}$} \\
demo + bia & -0.317 & 0.689 & 0.344 & 0.332 & 0.345 & 0.338 & 0.800 & 0.790 & 0.687 \\
\bottomrule
\end{tabular}%
}
\begin{flushleft} 
\vspace{0ex} 
{\footnotesize \textbf{*: statistical significance ($p < 0.05$)}.}
\end{flushleft}
\end{table}

\begin{table}[tbp]
\centering
\captionsetup{justification=centering}
\caption{Feature Importance Analysis Across Individual and Combined Feature Sets}
\label{tab:feature_importance}
\resizebox{\textwidth}{!}{%
\begin{tabular}{l|ccc|ccccc|ccc|ccc|c}
\toprule
\multirow{2}{*}{\textbf{Feature Set}} & \multicolumn{3}{c|}{\textbf{Demographics}} & \multicolumn{5}{c|}{\textbf{Anthropometrics}} & \multicolumn{3}{c|}{\textbf{DXA}} & \multicolumn{3}{c|}{\textbf{PhotoScan}} & \textbf{BIA} \\
 & \textbf{Age} & \textbf{BMI} & \textbf{Sex} & \textbf{Waist} & \textbf{Hip} & \textbf{WHR} & \textbf{WHtR} & \textbf{ABSI} & \textbf{Fat\%} & \textbf{A/G} & \textbf{V/S} & \textbf{Fat\%} & \textbf{A/G} & \textbf{V/S} & \textbf{Fat\%} \\
\midrule
demo & 0.215 & 0.769 & 0.016 & - & - & - & - & - & - & - & - & - & - & - & - \\
circ & - & - & - & 0.477 & 0.113 & 0.060 & 0.281 & 0.070 & - & - & - & - & - & - & - \\
dxa & - & - & - & - & - & - & - & - & 0.383 & 0.569 & 0.048 & - & - & - & - \\
photoscan & - & - & - & - & - & - & - & - & - & - & - & 0.243 & 0.595 & 0.162 & - \\
bia & - & - & - & - & - & - & - & - & - & - & - & - & - & - & 1.000 \\
\midrule
demo + circ & 0.078 & - & 0.000 & 0.461 & 0.102 & 0.046 & 0.256 & 0.056 & - & - & - & - & - & - & - \\
demo + dxa & 0.072 & 0.440 & 0.001 & - & - & - & - & - & 0.107 & 0.356 & 0.025 & - & - & - & - \\
demo + photoscan & 0.088 & 0.450 & 0.000 & - & - & - & - & - & - & - & - & 0.035 & 0.328 & 0.099 & - \\
circ + photoscan & - & - & - & 0.329 & 0.098 & 0.029 & 0.201 & 0.027 & - & - & - & 0.026 & 0.216 & 0.073 & - \\
demo + circ + photoscan & 0.056 & - & 0.000 & 0.323 & 0.093 & 0.023 & 0.192 & 0.022 & - & - & - & 0.021 & 0.203 & 0.067 & - \\
demo + bia & 0.183 & 0.698 & 0.003 & - & - & - & - & - & - & - & - & - & - & - & 0.116 \\
\bottomrule
\end{tabular}%
}
\end{table}

%% file: sections/3-discussion.tex
\section{Discussion}
\label{sec:discussion}

This study presents a comprehensive analysis of body composition and cardiometabolic health metrics, derived from a longitudinal MetabolicMosaic Cohort study. On two large studies (N=677), our proposed models on predicting total BF\% showed that PhotoScan, which includes BMI and sex, achieves state-of-the-art accuracy in predicting body fat percentage with a MAE of 2.15, compared to MAE=3.45 for demographics only (BMI, sex, and age), and MAE=2.84 when using BIA from wearable device. Fusing the predicted BF\% from PhotoScan with BIA further improves prediction accuracy, reducing MAE to 1.97. These results have been reproduced on an independent validation cohort, which shows the generalizability of the trained models to estimate BF\% from PhotoScan and BIA, and the fused combination of both.

Furthermore, our analysis on the estimation of cardiometabolic health using body composition measures illustrated that waist circumference, PhotoScan-predicted BF\%, and PhotoScan-predicted A/G ratio significantly enhance the accuracy of predicting insulin resistance compared to using demographics alone (\Cref{fig:feature_importance} and \Cref{tab:feature_importance}). In addition, BIA-derived BF\% does not add significant value towards predicting insulin resistance. Strong correlations are found between body composition metrics (waist circumference, hip circumference, BF\%, A/G ratio) and HOMA-IR (insulin resistance marker) as shown in \Cref{fig:mosaic_inference}H. Differences in fat storage patterns and associated risks between males and females were also observed and quantified.

The ubiquity of smartphones has made photo-based techniques for assessing body fat an extremely valuable scientific objective. Using a combination of PhotoScan and BIA, we are able to predict body fat percentage with a mean absolute error relative to DXA of 1.97 in the PhotoBIA cohort and 2.03 in the independent validation cohort (MetabolicMosaic Cohort), with no significant proportional bias. These results compare well with the previous literature in this area. ~\citep{Graybeal2023} investigated how Digital Imaging Analysis (DIA) using consumer smartphones could be used to estimate body fat percentage, fat mass, and fat-free mass. They studied the performance of a DIA application with developmental software trained by a 4-compartment model, on 102 healthy participants, using the camera of Apple iPhone and Samsung Galaxy consumer smartphones. They found Root Mean Square of $\sim$5\% for body fat percentage, with a large proportional bias, i.e. the predictions had a tendency to underestimate body fat percentage at large values and overestimate body fat percentage for small values. ~\citep{Majmudar2022} studied how consumer Apple iPhone smartphone cameras, as well as BIA devices could be used to estimate body fat percentage in an experiment with 134 healthy adults. For the camera images, participants were dressed in minimal, form-fitting clothing. The authors found that when compared against DXA measurements, the consumer smartphone images together with a convolutional neural network algorithm were able to predict body fat percentage with an MAE of 2.16 (1.88 for males, 2.34 for females), with no significant proportional bias. ~\citep{Ferreira2025} conducted a large study of 1,273 individuals to study whether photographs taken by consumer smartphones along with AI algorithms (AI-2D) could estimate body fat percentage. Imaging was done with participants dressed in bathing clothes using Apple iPhones. The authors found that images captured by consumer smartphones could predict body fat percentage with a MAE of 1.56, and no significant proportional bias, when compared to ground truth DXA scans, however there was no validation of this model on an independent cohort.

For predicting insulin resistance using body composition, we obtain an AUROC of 0.773 when using demographics along with body composition features derived from DXA. The classification AUROC is slightly lower (0.760) when we are limited to PhotoScan + demographics instead of using DXA. These results compare well with the existing literature. In a study of 140 healthy young Indonesian males, ~\citep{Kurniawan2018} found that body weight alone was a strong predictor of insulin resistance (they used a cutoff of 3.75 for HOMA-IR, which is more stringent than our cutoff of 2.9). They found that while several parameters are predictive of insulin resistance, body weight provided the highest AUROC (0.788), whereas waist circumference provided the high specificity (0.781) and body fat percentage showed the highest sensitivity (0.743). ~\citep{Cheng2017} conducted a study of 394 middle-aged and elderly Taiwanese individuals, to investigate the effect of metabolic parameters on insulin resistance (they used a cutoff of 2.3 for HOMA-IR which is less stringent than our choice of 2.9). They found that both BMI (AUROC = 0.749) and waist circumference (AUROC = 0.745) could predict insulin resistance well. Interestingly body fat percentage was less predictive than BMI (AUROC = 0.687). ~\citep{Racette2006} performed an investigation of 407 individuals aged 50–95 years to investigate predictors of insulin resistance (they used the insulin sensitivity index as a predictor of insulin resistance). They found that while several metabolic parameters are predictive of insulin resistance, waist circumference was the strongest predictor. Among obese children and adolescents, insulin resistance has been shown to be significantly associated with the A/G ratio ~\citep{Aucouturier2009}.

Furthermore, the distribution of an individual’s body fat or even their body shape may be a much more powerful indicator of their metabolic health status ~\citep{Okosun2015,Lee2025,Bai2025}. In an analysis of different measures to quantify body shape, Lee and colleagues were able to demonstrate that inclusion of waist circumference via a body shape index (ABSI) and waist-to-BMI ratio, had a stronger predictive power of CVD ~\citep{Lee2025} than BMI alone. Further, in individuals who already have poor cardiovascular health, the addition of ABSI outperforms body fat percentage when predicting adverse health outcomes, such as heart failure ~\citep{Bai2025}. These trends hold true when using android (trunk) to gynoid (hip) fat ratios, another way of quantifying body shape and fat distribution using DXA. In particular, a higher A/G ratio, has shown associations with elevated glucose, blood pressure, triglycerides and insulin resistance even in people within a healthy weight range ~\citep{Okosun2015}.

Although we used large cohorts to develop the PhotoScan technology (35,000 participants for the UKBB, 677 participants for PhotoBIA cohort), we acknowledge that the sample size of the MetabolicMosaic Cohort study and also the percent of subjects who underwent significant or notable changes in their body composition and overall health risk profile is a limitation of our study. While our recruitment was targeted towards subjects that were highly motivated to improve their health, in reality only a small subset experienced noteworthy changes (10 out of 132 subjects experienced 3\% body weight loss). Another limitation is the absence of more frequent laboratory measurements (for insulin resistance and body composition) to track changes in these metrics at a more granular level. Participants might have improved during the first two months and then regressed again towards the end of the study. Furthermore, more frequent measurements of BIA and PhotoScan at home could have provided additional evidence about the frequency of measurements that are required with these technologies in order to allow more accurate tracking of true body composition changes. The study design relies on a self-selected, non-controlled, external, paid lifestyle program. Since there was no standardized intervention, the specific nutritional and exercise programs varied greatly across participants, making it difficult to attribute changes to particular aspects of the lifestyle ~\citep{Okosun2015} modification. Moreover, depending on self-reported activity levels, nutrition habits, and intervention adherence can introduce inaccuracies and biases. To partially address recall bias, we incorporated a wearable tracker to passively monitor participants’ activity levels. It is worth pointing out a key limitation in existing studies, mainly the fact that their cross-sectional nature limits their ability to definitively establish causality between body composition and metabolic disease. Our study, while still limited in size, is a step towards a longitudinal study that attempts to examine the association between changes in body composition and fat distribution and metabolic health risk.

Future research should focus on leveraging our proposed model to create a more comprehensive digital phenotype from PhotoScans. A key extension would be to train the model to predict a full range of body circumference measurements (e.g., waist, hip, and waist-to-hip ratio). This would democratize the collection of anthropometric data, which is crucial for improving the prediction of insulin resistance. Moreover, the PhotoScan model could be expanded to estimate fundamental body composition metrics, such as the percentages of visceral fat, lean muscle mass, and total body water, that are established for their utility in refining cardiometabolic health assessment ~\citep{Okosun2015,Amato2010,BosyWestphal2018,Bell2018,Kaess2012}. In parallel, the predictive power of our models for insulin resistance could be substantially augmented through multi-modal data integration. Fusing our PhotoScan-derived features with longitudinal data from wearable sensors (e.g., activity and sleep patterns) ~\citep{Metwally2026}, precise glycemic variability captured by continuous glucose monitoring ~\citep{Metwally2024,Wu2025,Park2025,Klonoff2025}, and readily available routine blood biomarkers ~\citep{Metwally2026} represents a critical next step toward developing a holistic and highly accurate at-home risk prediction framework.

%% file: sections/4-methods.tex
\section{Methods}
    \label{sec:methods}
    
    \subsection{Study Design}
    We designed a 30-week, prospective longitudinal study to investigate the impact of self-selected, non-controlled diet and exercise intervention on various metrics related to cardiometabolic health under the approval of an Institutional Review Board (IRB). The study aimed to recruit over 150 individuals with normal to above-normal BMI from the San Francisco Bay Area. Participants were required to enroll in an external, paid lifestyle program and commit to remaining in the program for the entire study duration. To gauge program adherence, participants self-reported their monthly physical activity and dietary patterns, with a third-party vendor externally monitoring their engagement with the intervention. Participants were advised to be under a doctor’s care for any health-related interpretations.
    
    Participants were eligible if they were at least 18 years of age, reside in California, capable of standing and walking unaided, enrolled in a paid lifestyle improvement program, and willing to comply with study procedures and expectations. Key exclusion criteria included having medical conditions such as Type 1 diabetes or uncontrolled Type 2 diabetes (HbA1c $> 8.0\%$), severe hypertension (SBP $\ge 160$ or DBP $\ge 100$), or heart failure, and being pregnant. Additionally, individuals taking medications that affect heart rate, for diabetes, or to control glucose levels, as well as those with recent changes in medications for hypertension, diabetes, or heart failure, or weight exceeding 450lbs, were excluded. The study was approved by approved by WCG (Institutional Review Board \#1371945).
    
    The recruited subjects are required to: (1) Provide written informed consent, acknowledging that their de-identified data would be used for research purposes. (2) Attend in-person appointments at the Google Human Research Laboratory in San Francisco, for anthropometric measurements, blood pressure measurements, blood test, and body composition measurements at the start and end of the study. (3) Self-report their dietary patterns, physical activity and medications via questionnaires every month. (4) Wear a Fitbit Charge 6 for the passive collection of various physiological measures, including activity level, sleep and heart rate derived metrics for a minimum of 20 hours per day throughout the study. This was crucial for understanding how these metrics responded to lifestyle changes and correlated with other health indicators.
    
    \subsection{Cohorts and Data collection}
    
    \subsubsection{MetabolicMosaic Cohort study and data collection}
    At both the start and end of the 30-week study, participants completed a comprehensive battery of assessments. This began with a questionnaire detailing their demographics (race, ethnicity, sex), health history, medication use, exercise, smoking, and nutrition habits. Participants visited the Human Research Laboratory in San Francisco, for a series of physiological measurements.
    These included a DXA scan, BIA, PhotoScan images taken with Pixel phone, and anthropometrics measurements such as weight, height, waist and hip circumferences. Blood pressure was also recorded. A finger-stick blood sample was also drawn during these visits, requiring participants to fast for 12 hours prior, to measure fasting blood glucose, HgA1c, hemoglobin, and a lipid panel, including total cholesterol, LDL, VLDL, HDL and triglycerides. Additionally, a phlebotomist drew venous blood to measure fasting insulin and liver function, including total bilirubin, direct bilirubin, indirect bilirubin, alkaline phosphatase, and aspartate transaminase.
    
    As part of the study, Fitbit wearable data from a Charge 6 was also passively collected from participants continuously for at least 20 hours per day for the entire study duration. Their activity level (such as steps, active zone minutes), sleep metrics, and heart rate-derived measurements (such as RHR, HRV) were measured.
    
    The precision of the DXA machine in the Human Research Lab was determined by calculating RMS SD, Coefficient of Variation (CV), \%CV, and least significant change (LSC) for segmental and whole-body fat and lean mass. Interrater reliability between DXA technicians was assessed using mean difference $\pm95\%$ limits of agreement (LoA) with Bland–Altman plots and intraclass correlation coefficient (ICC). All statistical analyses were performed using Python, with a P value of $\le 0.05$ considered significant.
    
    The subset dataset consists of 195 visits contributed by 132 unique participants (110 first visits and 85 last visits). This specific cohort includes only participants with complete data collection across all three body composition modalities: DXA, PhotoScan, and BIA. The cohort’s key characteristics are summarized by the following means $\pm$ standard deviations: age $43.4 \pm 11.4$ years, BMI $29.0 \pm 7.1$, HOMA-IR $2.1 \pm 1.5$, diastolic blood pressure “DBP” $76.5 \pm 8.2$ mmHg, and systolic blood pressure “SBP” $117.4 \pm 14.3$ mmHg.
    
    \subsubsection{UKBB cohort and data collection}
    This research has been conducted using the UK Biobank Resource under Application Number 65275. The subset of the dataset containing both MRI images and DXA fat\% ground truth includes approximately 35,000 participants, with 18,000 females and 17,000 males. The age range spans from 45 to 82, with a mean age of 63.8 years. BF\% ranges from 9\% to 64\%, with an overall mean of 35\%. Male participants have a mean BF\% of 30.4\%, while female participants have a mean BF\% of 39.3\%. The dataset was randomly split into an 80\% training set and a 20\% evaluation set.
    
    \subsubsection{PhotoBIA cohort and data collection}
    PhotoBIA study was approved under IRB protocol Pro00065782 and was executed with the intention of collecting development data for body composition estimation algorithms utilizing PhotoScan and BIA. It was performed in two phases (PhotoBIA\_1, and PhotoBIA\_2). PhotoBIA\_1 included an initial cohort and PhotoBIA\_2 was designed in order to a) provide a more diverse set of subjects with smaller correlations between BMI and total body fat\% with an aim of specifically recruiting more muscular subjects and b) to provide more challenging cases for PhotoScan (with different types of clothes). \Cref{fig:bmi_vs_bodyfat} shows the correlation between BMI and BF\% for males and females in the two study cohorts, illustrating an expected smaller correlation in PhotoBIA\_2 even though that is also partly attributable to the more limited dynamic range of both variables. For the purpose of these studies subjects were asked to not eat or drink anything four hours in advance of their lab appointments in order to reduce the potential impact on BIA and DXA measurements.
    
    \subsection{Predicting Body Composition}
    
    \subsubsection{PhotoScan pretraining and fine tuning}
    The model is first pre-trained using UKBiobank MRI data for creating the images, and UKBB DXA derived fat\% as ground truth. The cohort containing both MRI and DXA is about 35k participants.
    
    We followed the methodology described in 42 to read the MRI data, the collection for a participant is split into six body segments, each containing slice data for the four MRI phases. We first combine the phases by summation, then we extract the exterior contour of the slice using OpenCV2 libraries. To remove spurious contours, we keep the largest contour, or the largest two if they have about the same area. Each slice has a z coordinate (representing the offset from the top of the scan), and contour points (x, y), giving us a space representation of the contour with (x, y, z) points. We stack all the slices to obtain a point cloud representation of the participant. To capture detailed shading and body structure, this mesh was rendered from multiple virtual camera positions, incorporating varied rotations and lighting conditions. Finally, 2D projections representing frontal and lateral (side) views were generated and combined into 300x300 pixel images, each containing one frontal and one side view.
    
    Following image processing, a standard ResNet-50 deep neural network (initialized with Imagenet weights) was trained. The network inputs consisted of the generated frontal and side view projections, along with participant sex, height, and weight (BMI was internally computed in the model). The head of the network combines the output of the ResNet50 with sex and bmi before a final dense layer with 64 units. Softmax is applied to produce the probability density function representing the fat\%. The dataset was partitioned into an 80\% training set and a 20\% test set. This pretrained model achieved a MAE of 2.01 for BF\% prediction on the UKBB test dataset.
    
    We then proceeded to finetune this pretrained model, originally developed with UKBB 3D mesh projections, using data from the PhotoBIA cohort. These cohorts provided mobile phone PhotoScans (front and side pictures) with corresponding DXA-derived BF\% as the ground truth (\Cref{tab:photobia_characteristics}). A critical step in this phase was to transform the RGB images from the PhotoScans to emulate the appearance of the UKBB mesh renderings. For this, we employed a landmark detection model, an internal version of the pose detection ML Kit 43 to facilitate body segmentation and the localization of anatomical joints. Images were cropped from the mouth to mid-knee and ankle levels, consistent with the UKBB image processing. The cropped front and side images were then packed (1 front + 1 left or right side) and resized to 300x300 pixels, and converted to grayscale. To enhance model robustness and generalize performance, the training dataset was augmented using random cropping to generate slight variation in the body location within the image, horizontal flipping, and random cutout techniques 44. Furthermore, the training set was supplemented with frames extracted from videos of participants completing a 360-degree rotation. A pipeline utilizing the aforementioned landmark model was employed to detect when a participant was facing the camera (frontal view) or exposing their right or left side (side view). This process was used to automatically select optimal frontal and lateral frames from these video sequences.
    
    Fine tuning was conducted using a 5-fold cross-validation strategy, ensuring a balanced distribution of BF\% values within each fold. The fine-tuned models demonstrated an average MAE of 2.2 (SD = 0.1) across all folds.
    
    \subsubsection{BIA hardware, pre training and fine tuning}
    Fitbit Sense 2 was used to measure BIA on the wrist. It implements a 4-wire sensing mechanism using a 50kHz excitation frequency and utilizes two bottom electrodes, two top electrodes, and TI’s AFE4500 chip. Extracting the value of the body’s impedance involves analyzing a high-dimensionality impedance network with 13 unknown parameters: 8 parasitic impedances, 4 contact impedances (ZE) and the desired body impedance (Zbody). The implemented solution relies on a divide-and-conquer approach: we first estimate 8 parasitic impedances during open-loop calibration, and then estimate 5 body-related impedances when the user is making contact with all electrodes. The following figure illustrates all the steps involved in estimating body composition parameters using BIA: a) signal acquisition using the hardware sensor and solving for Zbody, b) making a point estimate for Zbody using the BIA time-series and finally c) mapping BIA and demographics to body composition values.
    
    One significant challenge with the measurement setup is the utilization of dry electrodes with small contact area, which leads to large contact impedances which in turn can induce large errors when estimating the body’s impedance. Through our testing we determined that contact impedances up to 10kOhm are acceptable and lead to Zbody estimates with a mean absolute percentage error (MAPE) of 5\% or less (when compared to a ground truth BIA measurement obtained via a RJL Quantum IV or an Impedimed SFB7 used in the same posture), which is deemed as acceptable for accurate body fat percent estimation. In addition to that, in order to guarantee high precision in BIA measurements the subject’s posture needs to be controlled so that they achieve an armpit angle of around 90 degrees between the upper arm and the torso and also utilizing a hand gesture that prevents right and left hands/limbs from touching each other.
    
    Body composition models were initially developed and tested via cross validation in the PhotoBIA study cohorts. These models utilize both demographics (age, sex, height and weight) as inputs as well as the extracted body impedance (in the typical form of height squared over body resistance).
    
    In order to understand the impact of contact impedance in body impedance estimation, we also measured body impedance using a commercially available body impedance analyzer (RJL Quantum IV from RJL Systems) that has been widely used in research studies. Our analysis revealed that contact impedances as high as about 10k Ohm allow for acceptable errors, i.e. in the range of 5\% in terms of mean absolute error. That level of imprecision is deemed acceptable as it results in less than 1\% variation in the estimated body fat percent. A limitation of that threshold though is that as we saw in our studies, about one in three subjects cannot get a valid BIA-based body fat estimate during their first attempt, but that number drops to one in five subjects when subjects are allowed to take three successive measurements. This effect is also bound to be naturally mitigated even further in real life as subjects wear the devices continuously and sweat deposits build up on the electrodes which results in additional decrease in contact impedance and thus measurement availability.
    
    \subsection{Prediction of insulin resistance}
    
    \subsubsection{Inference of Body Composition in MetabolicMosaic Cohort Study using PhotoScan and BIA}
    
    \paragraph{Inference of Body Composition using PhotoScan}
    The MetabolicMosaic Cohort study collected participants' images at first and last visits, approximately 6 to 7 months apart. For each visit, images were captured from two views (front plus left and right side), two camera heights (phone on the floor, and phone at table height) and two poses (“A pause” and “Relaxed”). Of the 140 participants, 911 images were generated for model inference. We averaged the inference results of the left and right side images (for a given pose and phone height). Notably one participant was completely excluded due to being near the DXA weight limit, and another participant had four side images excluded because of errors with the image capture (missing front picture).
    
    \paragraph{Inference of Body Composition using BIA}
    Sense 2 BIA data were collected from the study’s participants under resting conditions while the subjects were seated, with their elbows raised and while avoiding “cross touch”, i.e. contact between the two hands. Two collections were made for each participant and in the final estimation the average of the two BIA values were if both were available, otherwise only one measurement was used if the other one was not marked as valid. A body fat prediction model trained and tuned on the PhotoBIA cohort was used for inference and it was compared with a similarly trained demographic model. The following figure illustrates that the model achieved reasonably good performance and in line with expectations based on previously estimated performance on our previous studies. The model was able to achieve a mean absolute error of 2.66\% and a Pearson correlation of 0.9 compared to 3.39\% and 0.85 respectively from the demographic based model.
    
    \subsubsection{Data preprocessing}
    \begin{itemize}
        \item \textbf{Demographics:} We extracted users’ age and sex at birth from the onboarding survey, the BMI is computed from the on-site measurement of height and weight.
        \item \textbf{PhotoScan:} For each visit, users took multiple photos at two camera positions (phone on the floor, and phone at table height.) and two poses, “A pause” and “Relaxed”. We averaged the inference results of the left and right sides (for a given pose and phone position). One user’s last visit has swapped front/ side photos, hence was excluded and has no PhotoScan derived body composition.
        \item \textbf{DXA:} Each subject takes one DXA scan at the first and one scan at the last visit, only one subject was excluded due to a suspicious lower leg swelling which affects the accuracy of fat mass measurement.
        \item \textbf{BIA:} Each subject takes two consecutive BIA measurements for each visit, and the aggregation of the BF\% estimations from these BIA measurements are used.
        \item \textbf{Data Standardization:} The data used for modeling is a concatenation of demographics, anthropometrics, and body compositions from DXA scan, PhotoScan and BIA. In order to create a consistent modeling data that is agnostic to the learning model, we standardized input features to have zero mean and unit variance. For each training fold, our “normalizer” object was fit to the data in the training subset (not including samples in the testing subset). The fitted object was then used to transform the samples in both the training and testing subsets. The standardized data was used for all modeling tasks and evaluations.
    \end{itemize}
    
    \subsubsection{Modeling}
    Using the MetabolicMosaic Cohort, we trained one Gradient Boosting classifier to predict insulin resistance (vs. non-insulin resistance).
    
    \textbf{Feature Set Combination:} Overall there are 5 feature sets, demographics, anthropometrics, body compositions from DXA, body compositions from PhotoScan, body compositions from BIA. To decide which feature set or feature sets’ combination performs the best, our experiments trained each feature set and their combinations in parallel, for later comparison.
    
    \textbf{Removal of Collinear Features:} Given that XGBoost is generally robust to multicollinearity, we applied a high correlation threshold to eliminate the near-duplicate features. We assessed Pearson correlation coefficients for all feature pairs; for any pair exceeding 0.95 , we retained the feature with the stronger correlation to HOMA-IR and removed the other.
    
    \textbf{Repeated Nested Cross-Validation:}
    Given the limited dataset (N=195), relying on a simple hold-out test set would likely yield high variance. Therefore, to ensure robust and unbiased performance estimation, we employed a Repeated Nested Cross-Validation framework. This procedure separates model selection from evaluation to prevent data leakage:
    \begin{itemize}
        \item \textbf{Inner Loop (Model Selection):} A 3-fold stratified grid search was performed on the training portion of each split to optimize hyperparameters. The parameter grid is detailed below:
        \begin{itemize}
            \item learning\_rate: [0.01, 0.02, 0.05]
            \item n\_estimators: [100, 200, 300]
            \item min\_samples\_split: [20, 40, 80]
            \item min\_samples\_leaf: [5, 10, 20]
            \item subsample: [0.8, 1]
            \item pos\_sample\_weight: [2, 5]
            \item max\_depth: [3, 5]
        \end{itemize}
        \item \textbf{Outer Loop (Performance Evaluation):} A Repeated Stratified K-Fold procedure (10 folds, 5 repeats) was applied to assess generalization.
    \end{itemize}
    To maximize statistical power while maintaining test-set independence, final performance metrics (e.g., DeLong test p-value, NRI) were calculated based on aggregated consensus probabilities . These were derived by averaging the out-of-fold predictions for each subject across the 5 repetitions, ensuring a stable assessment of the full cohort.
    
    \subsubsection{Evaluation}
    Model performance was comprehensively assessed on each of the 50 test folds using AUROC, AUPRC and NRI 45. Additionally, NPV, Sensitivity, and Precision were evaluated at a fixed specificity of 80\%. Furthermore, the feature importance, calculated as the mean and standard deviation across the 50 training sets, is summarized in \Cref{fig:feature_importance} and \Cref{tab:feature_importance} for the respective classification models.
    
    \subsubsection{Statistical analysis}
    To statistically compare the performance of the selected feature sets, we employed DeLong’s test for correlated ROC curves. We aggregated the out-of-fold predicted probabilities from the repeated stratified k-fold cross-validation. For each subject, the final prediction was derived by averaging the probabilities from the validation repeats. These pooled predictions were used to generate the global ROC curves for each model. We conducted pairwise comparisons between each feature set and the Demographics-only baseline model. The resulting two-sided p-values are presented in \Cref{tab:model_performance}, with statistical significance defined as $p < 0.05$.
    
    \subsubsection{Visualization methods}
    We used Matplotlib, seaborn, plotly and Figma to plot most of the figures.

%% file: supplementary/supplementary_figures.tex
\label{supp:figures}

\setcounter{figure}{0}
\renewcommand{\thefigure}{S\arabic{figure}}

\begin{figure*}[htbp]
    \centering
    \includegraphics[width=0.95\textwidth]{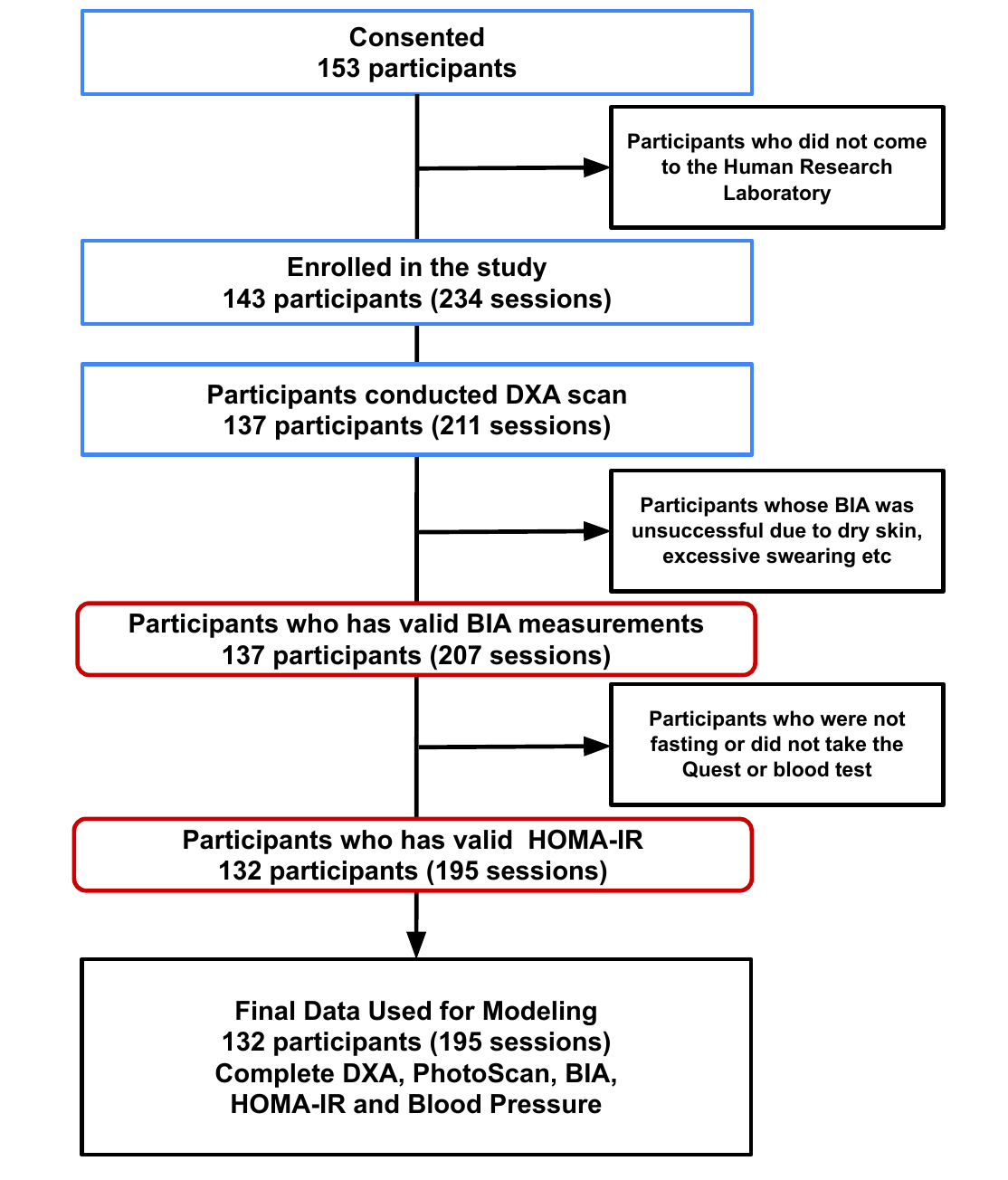}
    \caption{\textbf{Data Processing Workflow of the MetabolicMosaic Cohort Study.}}
    \label{fig:data_processing_workflow}
\end{figure*}

\begin{figure*}[htbp]
    \centering
    \includegraphics[width=0.95\textwidth]{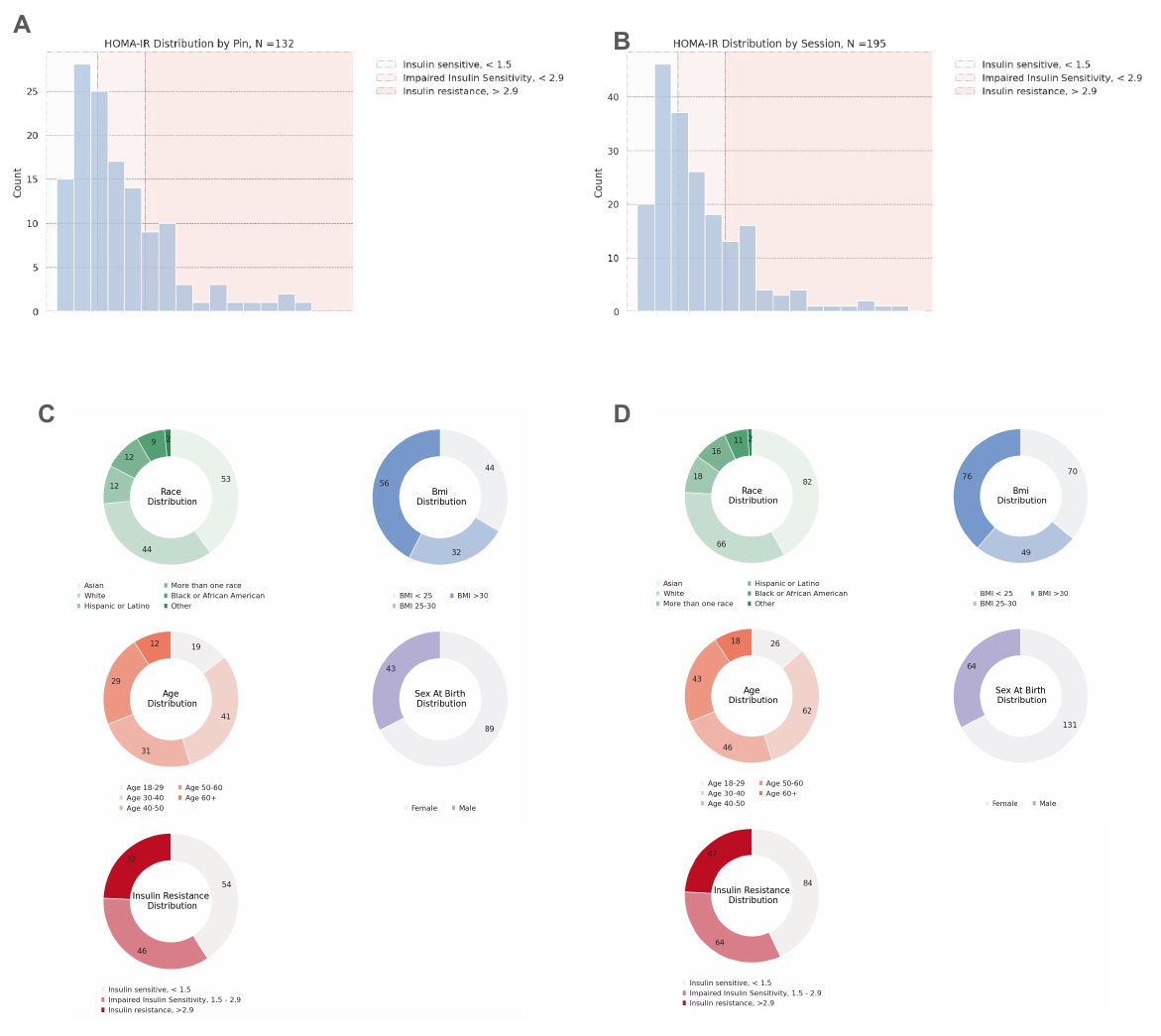}
    \caption{\textbf{MetabolicMosaic Cohort Characteristics.} A) HOMA-IR Histogram for all participants (first visits only). B) HOMA-IR Distribution for all visits. C) Demographic breakdown by race, BMI, age, sex, insulin resistance status and hypertension stages for all participants (first visits only). D) Demographic breakdown by race, BMI, age, sex, and insulin resistance status for all visits.}
    \label{fig:metabolicmosaic_cohort}
\end{figure*}

\begin{figure*}[htbp]
    \centering
    \includegraphics[width=0.75\textwidth]{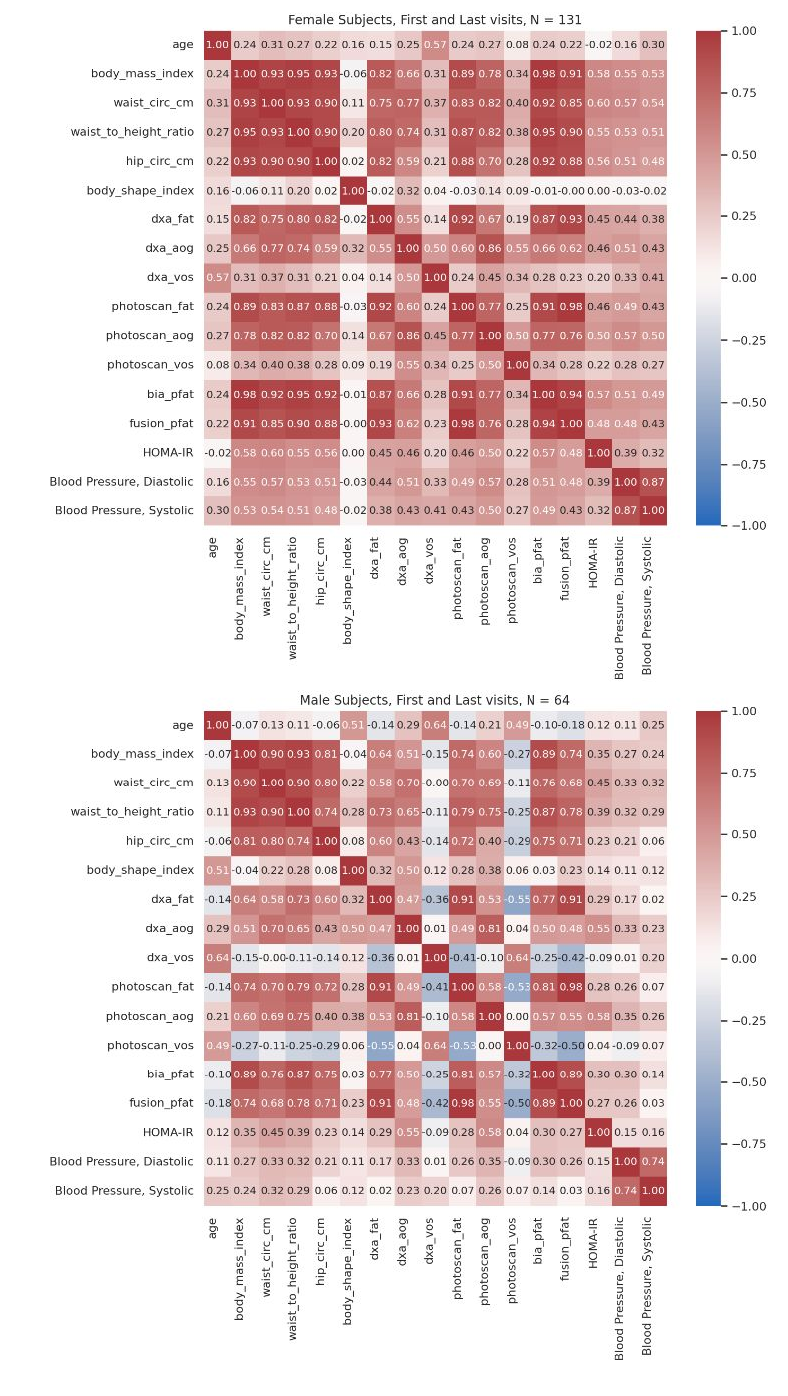}
    \caption{\textbf{Correlation between Body Composition metrics, Demographics, Anthropometrics, and Cardiometabolic Health Metrics with Gender Stratification.} A) All visits from Female Participants B) All visits from Male Participants.}
    \label{fig:feature_correlation_by_gender}
\end{figure*}

\begin{figure*}[htbp]
    \centering
    \includegraphics[width=0.95\textwidth]{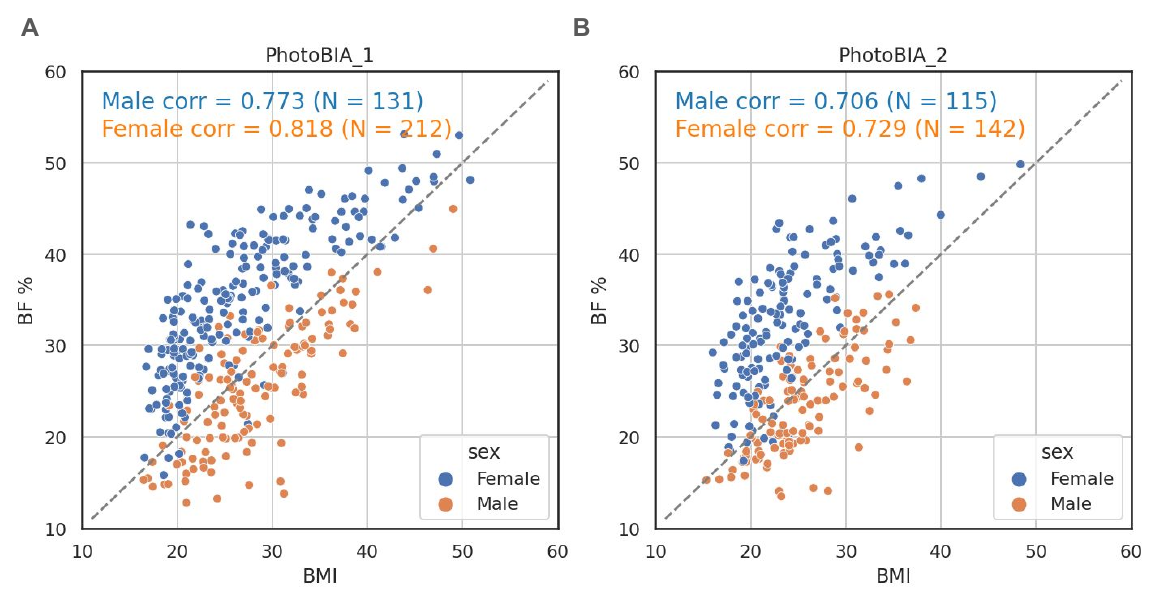}
    \caption{\textbf{Scatter Plots of BMI against DXA-derived BF\% in the PhotoBIA Cohorts.} A) PhotoBIA\_1 Cohort. B) PhotoBIA\_2 Cohort.}
    \label{fig:bmi_vs_bodyfat}
\end{figure*}

%% file: supplementary/supplementary_tables.tex
\label{supp:tables}
\setcounter{table}{0}
\renewcommand{\thetable}{S\arabic{table}}

\begin{table}[htbp]
\centering
\caption{MetabolicMosaic Study Cohort Characteristics by Visits for Insulin Resistance}
\label{tab:cohort_characteristics}
\resizebox{\textwidth}{!}{%
\begin{tabular}{ll|ccccc|ccccc|ccccc}
\toprule
 & & \multicolumn{5}{c|}{\textbf{First Visits}} & \multicolumn{5}{c|}{\textbf{Last Visits}} & \multicolumn{5}{c}{\textbf{Subjects with Both Visits}} \\
 & & \textbf{IS} & \textbf{Impaired} & \textbf{IR} & \textbf{All} & \textbf{P} & \textbf{IS} & \textbf{Impaired} & \textbf{IR} & \textbf{All} & \textbf{P} & \textbf{IS} & \textbf{Impaired} & \textbf{IR} & \textbf{All} & \textbf{P} \\
\midrule
\multicolumn{2}{l|}{\textbf{Sample Size}} & 43 & 41 & 26 & 110 & - & 41 & 23 & 21 & 85 & - & 25 & 24 & 14 & 63 & - \\
\multicolumn{2}{l|}{\textbf{\# of Sessions}} & 43 & 41 & 26 & 110 & - & 41 & 23 & 21 & 85 & - & 43 & 54 & 29 & 126 & - \\
\midrule
\multirow{2}{*}{\textbf{Sex}} & Female & 28 & 28 & 16 & 72 & 0.85 & 26 & 18 & 15 & 59 & 0.45 & 33 & 36 & 15 & 84 & 0.09 \\
 & Male & 15 & 13 & 10 & 38 & 0.85 & 15 & 5 & 6 & 26 & 0.45 & 10 & 18 & 14 & 42 & 0.09 \\
\midrule
\multirow{5}{*}{\textbf{Ethnicity}} & Asian & 17 & 20 & 8 & 45 & 0.16 & 19 & 8 & 10 & 37 & 0.08 & 18 & 25 & 15 & 58 & 0.13 \\
 & Black/African Am. & 1 & 2 & 4 & 7 & 0.16 & 1 & 1 & 2 & 4 & 0.08 & 1 & 1 & 2 & 4 & 0.13 \\
 & Hispanic & 2 & 3 & 5 & 10 & 0.16 & 0 & 2 & 4 & 6 & 0.08 & 4 & 0 & 4 & 8 & 0.13 \\
 & White & 17 & 12 & 8 & 37 & 0.16 & 16 & 9 & 4 & 29 & 0.08 & 14 & 23 & 7 & 44 & 0.13 \\
 & Other/Mixed & 6 & 4 & 1 & 11 & 0.16 & 5 & 3 & 1 & 9 & 0.08 & 6 & 5 & 1 & 12 & 0.13 \\
\midrule
\textbf{Demog.} & Age (years) & \makecell{43.4 \\ (10.8)} & \makecell{42.9 \\ (10.4)} & \makecell{44.0 \\ (12.7)} & \makecell{43.4 \\ (11.0)} & 0.93 & \makecell{42.8 \\ (12.0)} & \makecell{43.8 \\ (11.6)} & \makecell{44.7 \\ (12.7)} & \makecell{43.5 \\ (11.9)} & 0.83 & \makecell{44.2 \\ (11.0)} & \makecell{43.1 \\ (11.2)} & \makecell{45.3 \\ (13.6)} & \makecell{44.0 \\ (11.7)} & 0.69 \\
 & BMI ($kg/m^2$) & \makecell{26.0 \\ (6.4)} & \makecell{29.1 \\ (6.6)} & \makecell{34.9 \\ (7.8)} & \makecell{29.2 \\ (7.6)} & $<0.001$ & \makecell{26.1 \\ (4.3)} & \makecell{29.1 \\ (6.3)} & \makecell{33.1 \\ (7.5)} & \makecell{28.6 \\ (6.4)} & $<0.001$ & \makecell{27.9 \\ (6.1)} & \makecell{25.2 \\ (3.6)} & \makecell{32.2 \\ (6.9)} & \makecell{27.7 \\ (6.0)} & $<0.001$ \\
\midrule
\multirow{7}{*}{\textbf{Biomarkers}} & HDL & \makecell{56.3 \\ (12.4)} & \makecell{48.2 \\ (12.1)} & \makecell{41.7 \\ (14.5)} & \makecell{49.8 \\ (13.9)} & $<0.001$ & \makecell{59.5 \\ (18.2)} & \makecell{53.3 \\ (14.8)} & \makecell{43.8 \\ (15.1)} & \makecell{53.9 \\ (17.6)} & 0.004 & \makecell{50.7 \\ (12.6)} & \makecell{58.2 \\ (16.1)} & \makecell{42.4 \\ (15.8)} & \makecell{52.0 \\ (16.0)} & $<0.001$ \\
 & HbA1c & \makecell{4.9 \\ (0.4)} & \makecell{5.0 \\ (0.3)} & \makecell{5.1 \\ (0.5)} & \makecell{5.0 \\ (0.4)} & 0.12 & \makecell{5.2 \\ (0.9)} & \makecell{5.3 \\ (0.8)} & \makecell{5.5 \\ (1.0)} & \makecell{5.3 \\ (0.9)} & 0.58 & \makecell{5.1 \\ (0.6)} & \makecell{5.1 \\ (0.8)} & \makecell{5.4 \\ (0.9)} & \makecell{5.2 \\ (0.8)} & 0.31 \\
 & HOMA-IR & \makecell{1.0 \\ (0.3)} & \makecell{2.0 \\ (0.4)} & \makecell{4.4 \\ (1.6)} & \makecell{2.2 \\ (1.5)} & $<0.001$ & \makecell{1.1 \\ (0.3)} & \makecell{2.1 \\ (0.4)} & \makecell{4.1 \\ (1.4)} & \makecell{2.1 \\ (1.4)} & $<0.001$ & \makecell{2.0 \\ (0.4)} & \makecell{1.0 \\ (0.3)} & \makecell{3.9 \\ (1.4)} & \makecell{2.0 \\ (1.4)} & $<0.001$ \\
 & LDL & \makecell{126.6 \\ (25.6)} & \makecell{119.2 \\ (36.6)} & \makecell{113.0 \\ (31.3)} & \makecell{120.4 \\ (31.6)} & 0.23 & \makecell{126.8 \\ (34.6)} & \makecell{134.2 \\ (34.2)} & \makecell{130.0 \\ (38.8)} & \makecell{129.8 \\ (35.3)} & 0.77 & \makecell{132.9 \\ (35.5)} & \makecell{129.4 \\ (31.1)} & \makecell{120.2 \\ (33.1)} & \makecell{128.2 \\ (33.3)} & 0.30 \\
 & TAG & \makecell{95.1 \\ (43.3)} & \makecell{107.6 \\ (52.5)} & \makecell{139.2 \\ (75.0)} & \makecell{110.8 \\ (58.1)} & 0.009 & \makecell{84.1 \\ (40.5)} & \makecell{103.8 \\ (53.8)} & \makecell{135.2 \\ (85.8)} & \makecell{103.8 \\ (62.7)} & 0.01 & \makecell{105.5 \\ (48.8)} & \makecell{90.2 \\ (46.9)} & \makecell{133.7 \\ (63.5)} & \makecell{106.7 \\ (54.6)} & 0.004 \\
 & FSBG & \makecell{89.8 \\ (8.3)} & \makecell{93.2 \\ (8.6)} & \makecell{101.4 \\ (12.9)} & \makecell{93.8 \\ (10.6)} & $<0.001$ & \makecell{87.8 \\ (7.3)} & \makecell{92.2 \\ (8.2)} & \makecell{100.0 \\ (12.3)} & \makecell{92.0 \\ (10.2)} & $<0.001$ & \makecell{93.9 \\ (8.3)} & \makecell{88.9 \\ (7.7)} & \makecell{101.2 \\ (13.0)} & \makecell{93.5 \\ (10.4)} & $<0.001$ \\
 & Total Chol. & \makecell{196.5 \\ (32.5)} & \makecell{187.2 \\ (40.6)} & \makecell{182.5 \\ (36.6)} & \makecell{189.8 \\ (36.8)} & 0.27 & \makecell{199.0 \\ (33.3)} & \makecell{191.0 \\ (57.3)} & \makecell{204.7 \\ (45.7)} & \makecell{198.2 \\ (43.7)} & 0.58 & \makecell{198.5 \\ (40.4)} & \makecell{199.2 \\ (32.9)} & \makecell{192.1 \\ (36.5)} & \makecell{197.3 \\ (36.3)} & 0.67 \\
\midrule
\multirow{5}{*}{\textbf{Body Comp}} & Waist (cm) & \makecell{83.4 \\ (14.9)} & \makecell{90.7 \\ (15.2)} & \makecell{104.0 \\ (15.0)} & \makecell{91.0 \\ (16.9)} & $<0.001$ & \makecell{84.0 \\ (9.7)} & \makecell{91.0 \\ (13.3)} & \makecell{102.3 \\ (14.0)} & \makecell{90.4 \\ (13.9)} & $<0.001$ & \makecell{88.1 \\ (13.8)} & \makecell{81.5 \\ (8.6)} & \makecell{100.5 \\ (13.5)} & \makecell{88.1 \\ (13.8)} & $<0.001$ \\
 & Hip (cm) & \makecell{99.3 \\ (12.9)} & \makecell{104.8 \\ (14.0)} & \makecell{115.4 \\ (16.6)} & \makecell{105.1 \\ (15.4)} & $<0.001$ & \makecell{103.2 \\ (8.9)} & \makecell{106.7 \\ (13.9)} & \makecell{117.8 \\ (16.2)} & \makecell{107.8 \\ (13.6)} & $<0.001$ & \makecell{103.2 \\ (12.1)} & \makecell{100.7 \\ (8.8)} & \makecell{112.3 \\ (14.1)} & \makecell{104.2 \\ (12.1)} & $<0.001$ \\
 & Body Fat \% & \makecell{32.5 \\ (7.5)} & \makecell{34.8 \\ (8.2)} & \makecell{38.3 \\ (8.5)} & \makecell{34.7 \\ (8.2)} & 0.02 & \makecell{31.2 \\ (7.7)} & \makecell{34.5 \\ (7.0)} & \makecell{38.6 \\ (8.3)} & \makecell{34.0 \\ (8.2)} & 0.002 & \makecell{34.2 \\ (7.1)} & \makecell{31.5 \\ (7.3)} & \makecell{36.4 \\ (8.4)} & \makecell{33.6 \\ (7.7)} & 0.02 \\
 & A/G & \makecell{0.4 \\ (0.2)} & \makecell{0.5 \\ (0.1)} & \makecell{0.6 \\ (0.2)} & \makecell{0.5 \\ (0.2)} & $<0.001$ & \makecell{0.4 \\ (0.1)} & \makecell{0.5 \\ (0.1)} & \makecell{0.6 \\ (0.2)} & \makecell{0.5 \\ (0.2)} & $<0.001$ & \makecell{0.5 \\ (0.1)} & \makecell{0.4 \\ (0.1)} & \makecell{0.6 \\ (0.2)} & \makecell{0.5 \\ (0.2)} & $<0.001$ \\
 & V/S & \makecell{0.3 \\ (0.1)} & \makecell{0.3 \\ (0.1)} & \makecell{0.3 \\ (0.1)} & \makecell{0.3 \\ (0.1)} & 0.48 & \makecell{0.3 \\ (0.2)} & \makecell{0.3 \\ (0.1)} & \makecell{0.3 \\ (0.1)} & \makecell{0.3 \\ (0.1)} & 0.85 & \makecell{0.3 \\ (0.1)} & \makecell{0.3 \\ (0.1)} & \makecell{0.3 \\ (0.1)} & \makecell{0.3 \\ (0.1)} & 0.31 \\
\midrule
\multirow{2}{*}{\textbf{Blood Pressure}} & Diastolic & \makecell{74.4 \\ (7.7)} & \makecell{76.9 \\ (7.6)} & \makecell{82.1 \\ (8.3)} & \makecell{77.2 \\ (8.3)} & $<0.001$ & \makecell{72.5 \\ (6.9)} & \makecell{78.3 \\ (7.8)} & \makecell{79.0 \\ (8.8)} & \makecell{75.7 \\ (8.2)} & 0.002 & \makecell{78.0 \\ (6.5)} & \makecell{72.5 \\ (6.7)} & \makecell{80.7 \\ (9.3)} & \makecell{76.3 \\ (8.0)} & $<0.001$ \\
 & Systolic & \makecell{114.4 \\ (14.6)} & \makecell{118.9 \\ (14.5)} & \makecell{128.2 \\ (14.3)} & \makecell{119.3 \\ (15.3)} & $<0.001$ & \makecell{112.0 \\ (11.6)} & \makecell{117.0 \\ (11.5)} & \makecell{118.0 \\ (14.8)} & \makecell{114.8 \\ (12.6)} & 0.13 & \makecell{118.5 \\ (12.2)} & \makecell{112.0 \\ (12.0)} & \makecell{123.6 \\ (17.0)} & \makecell{116.9 \\ (14.0)} & $<0.001$ \\
\bottomrule
\end{tabular}%
}
\end{table}

\begin{table}[htbp]
\centering
\caption{BF\% Estimation Accuracy of PhotoScan Stratified by Sex, Camera Position, Pose, and Ethnicity for MetabolicMosaic Cohort}
\label{tab:stratified_bf}
\resizebox{\textwidth}{!}{%
\begin{tabular}{l|c|cc|cc|cc|ccccc}
\toprule
 & \textbf{All} & \multicolumn{2}{c|}{\textbf{Sex}} & \multicolumn{2}{c|}{\textbf{Camera Position}} & \multicolumn{2}{c|}{\textbf{Pose}} & \multicolumn{5}{c}{\textbf{Ethnicity}} \\
 & & \textbf{M} & \textbf{F} & \textbf{Table} & \textbf{Floor} & \textbf{Relaxed} & \textbf{A-pose} & \textbf{Asian} & \textbf{Black} & \textbf{Hisp.} & \textbf{Other} & \textbf{White} \\
\midrule
\textbf{N} & 140 & 45 & 95 & 138 & 140 & 140 & 140 & 57 & 11 & 12 & 8 & 52 \\
\textbf{MAE} & 2.20 & 1.83 & 2.38 & 2.26 & 2.14 & 2.24 & 2.15 & 2.16 & 2.85 & 2.02 & 1.69 & 2.21 \\
\textbf{StdDev} & 1.74 & 1.77 & 1.70 & 1.72 & 1.75 & 1.75 & 1.73 & 1.73 & 1.70 & 1.56 & 1.44 & 1.81 \\
\textbf{RMSE} & 2.80 & 2.54 & 2.92 & 2.84 & 2.76 & 2.84 & 2.76 & 2.77 & 3.31 & 2.54 & 2.21 & 2.86 \\
\textbf{Mean BF\%} & 34.4 & 27.2 & 37.9 & 34.3 & 34.4 & 34.4 & 34.4 & 32.1 & 39.5 & 44.4 & 35.4 & 33.8 \\
\bottomrule
\end{tabular}%
}
\end{table}

\begin{table}[htbp]
\centering
\caption{Estimation Accuracy of A/G and V/S on PhotoBIA Cohort and MetabolicMosaic Cohort}
\label{tab:ag_vs_performance}
\begin{tabular}{lllccc}
\toprule
\textbf{Cohort} & \textbf{Metric} & \textbf{Method} & \textbf{All} & \textbf{Female} & \textbf{Male} \\
\midrule
\multirow{6}{*}{\textbf{PhotoBIA}} & \multirow{2}{*}{N} & & 677 & 385 & 289 \\
 & & & & & \\
 & \multirow{2}{*}{\textbf{A/G}} & Demographics  & 0.111 & 0.103 & 0.123 \\
 & & PhotoScan & \textbf{0.107} & \textbf{0.091} & 0.129 \\
 \cmidrule{2-6}
 & \multirow{2}{*}{\textbf{V/S}} & Demographics & 0.106 & 0.067 & 0.156 \\
 & & PhotoScan  & \textbf{0.094} & 0.078 & \textbf{0.116} \\
\midrule
\multirow{6}{*}{\textbf{MetabolicMosaic}} & \multirow{2}{*}{N (Sessions)} & & \makecell{140 \\ (231)} & \makecell{95 \\ (154)} & \makecell{45 \\ (77)} \\
 & & & & & \\
 & \multirow{2}{*}{\textbf{A/G}} & Demographics  & 0.116 & 0.099 & 0.152 \\
 & & PhotoScan  & \textbf{0.085} & \textbf{0.070} & \textbf{0.116} \\
 \cmidrule{2-6}
 & \multirow{2}{*}{\textbf{V/S}} & Demographics  & 0.101 & 0.078 & 0.148 \\
 & & PhotoScan  & \textbf{0.085} & \textbf{0.073} & \textbf{0.110} \\
\bottomrule
\end{tabular}%
\end{table}

\renewcommand{\thetable}{S4\alph{table}}
\setcounter{table}{0} 
\begin{sidewaystable}[p]
\centering
\captionsetup{justification=centering}
\caption{Feature Correlation of All Subjects}
\label{tab:feature_correlation_all_subjects_final}
\tiny
\setlength{\tabcolsep}{1pt}
\begin{tabular}{lcccccccccccccccccc}
\toprule
\textbf{Feature} & \textbf{age} & \makecell{\textbf{body\_mass\_}\\\textbf{index}} & \makecell{\textbf{waist\_}\\\textbf{circ\_cm}} & \makecell{\textbf{hip\_}\\\textbf{circ\_cm}} & \makecell{\textbf{waist\_to\_}\\\textbf{hip\_ratio}} & \makecell{\textbf{waist\_to\_}\\\textbf{height\_ratio}} & \makecell{\textbf{body\_shape\_}\\\textbf{index}} & \makecell{\textbf{dxa\_}\\\textbf{fat}} & \makecell{\textbf{dxa\_}\\\textbf{aog}} & \makecell{\textbf{dxa\_}\\\textbf{vos}} & \makecell{\textbf{photoscan\_}\\\textbf{fat}} & \makecell{\textbf{photoscan\_}\\\textbf{aog}} & \makecell{\textbf{photoscan\_}\\\textbf{vos}} & \makecell{\textbf{bia\_}\\\textbf{pfat}} & \makecell{\textbf{fusion\_}\\\textbf{pfat}} & \textbf{HOMA-IR} & \makecell{\textbf{BP}\\\textbf{Diastolic}} & \makecell{\textbf{BP}\\\textbf{Systolic}} \\
\midrule
BP Diastolic & \makecell{0.14 \\ (0.04)} & \makecell{0.45 \\ (<0.001)} & \makecell{0.51 \\ (<0.001)} & \makecell{0.41 \\ (<0.001)} & \makecell{0.25 \\ (<0.001)} & \makecell{0.45 \\ (<0.001)} & \makecell{0.04 \\ (0.57)} & \makecell{0.25 \\ (<0.001)} & \makecell{0.42 \\ (<0.001)} & \makecell{0.18 \\ (0.01)} & \makecell{0.27 \\ (<0.001)} & \makecell{0.46 \\ (<0.001)} & \makecell{0.11 \\ (0.13)} & \makecell{0.30 \\ (<0.001)} & \makecell{0.28 \\ (<0.001)} & \makecell{0.30 \\ (<0.001)} & -- & \makecell{0.82 \\ (<0.001)} \\ \addlinespace
BP Systolic & \makecell{0.29 \\ (<0.001)} & \makecell{0.40 \\ (<0.001)} & \makecell{0.48 \\ (<0.001)} & \makecell{0.31 \\ (<0.001)} & \makecell{0.35 \\ (<0.001)} & \makecell{0.40 \\ (<0.001)} & \makecell{0.09 \\ (0.23)} & \makecell{0.10 \\ (0.17)} & \makecell{0.40 \\ (<0.001)} & \makecell{0.35 \\ (<0.001)} & \makecell{0.11 \\ (0.12)} & \makecell{0.45 \\ (<0.001)} & \makecell{0.25 \\ (<0.001)} & \makecell{0.17 \\ (0.02)} & \makecell{0.12 \\ (0.08)} & \makecell{0.26 \\ (<0.001)} & \makecell{0.82 \\ (<0.001)} & -- \\ \addlinespace
HOMA-IR & \makecell{0.04 \\ (0.57)} & \makecell{0.48 \\ (<0.001)} & \makecell{0.54 \\ (<0.001)} & \makecell{0.43 \\ (<0.001)} & \makecell{0.28 \\ (<0.001)} & \makecell{0.47 \\ (<0.001)} & \makecell{0.06 \\ (0.39)} & \makecell{0.29 \\ (<0.001)} & \makecell{0.45 \\ (<0.001)} & \makecell{0.06 \\ (0.42)} & \makecell{0.27 \\ (<0.001)} & \makecell{0.49 \\ (<0.001)} & \makecell{0.10 \\ (0.15)} & \makecell{0.34 \\ (<0.001)} & \makecell{0.29 \\ (<0.001)} & -- & \makecell{0.30 \\ (<0.001)} & \makecell{0.26 \\ (<0.001)} \\ \addlinespace
age & -- & \makecell{0.13 \\ (0.06)} & \makecell{0.25 \\ (<0.001)} & \makecell{0.13 \\ (0.08)} & \makecell{0.22 \\ (0.002)} & \makecell{0.21 \\ (0.004)} & \makecell{0.29 \\ (<0.001)} & \makecell{0.01 \\ (0.86)} & \makecell{0.26 \\ (<0.001)} & \makecell{0.53 \\ (<0.001)} & \makecell{0.06 \\ (0.41)} & \makecell{0.24 \\ (0.001)} & \makecell{0.23 \\ (0.001)} & \makecell{0.07 \\ (0.36)} & \makecell{0.05 \\ (0.52)} & \makecell{0.04 \\ (0.57)} & \makecell{0.14 \\ (0.04)} & \makecell{0.29 \\ (<0.001)} \\ \addlinespace
bia\_pfat & \makecell{0.07 \\ (0.36)} & \makecell{0.84 \\ (<0.001)} & \makecell{0.61 \\ (<0.001)} & \makecell{0.80 \\ (<0.001)} & \makecell{-0.01 \\ (0.88)} & \makecell{0.82 \\ (<0.001)} & \makecell{-0.21 \\ (0.004)} & \makecell{0.90 \\ (<0.001)} & \makecell{0.10 \\ (0.15)} & \makecell{-0.30 \\ (<0.001)} & \makecell{0.93 \\ (<0.001)} & \makecell{0.19 \\ (0.008)} & \makecell{-0.46 \\ (<0.001)} & -- & \makecell{0.96 \\ (<0.001)} & \makecell{0.34 \\ (<0.001)} & \makecell{0.30 \\ (<0.001)} & \makecell{0.17 \\ (0.02)} \\ \addlinespace
body\_mass\_index & \makecell{0.13 \\ (0.06)} & -- & \makecell{0.88 \\ (<0.001)} & \makecell{0.91 \\ (<0.001)} & \makecell{0.30 \\ (<0.001)} & \makecell{0.95 \\ (<0.001)} & \makecell{-0.11 \\ (0.14)} & \makecell{0.72 \\ (<0.001)} & \makecell{0.43 \\ (<0.001)} & \makecell{0.01 \\ (0.88)} & \makecell{0.75 \\ (<0.001)} & \makecell{0.54 \\ (<0.001)} & \makecell{-0.08 \\ (0.29)} & \makecell{0.84 \\ (<0.001)} & \makecell{0.78 \\ (<0.001)} & \makecell{0.48 \\ (<0.001)} & \makecell{0.45 \\ (<0.001)} & \makecell{0.40 \\ (<0.001)} \\ \addlinespace
body\_shape\_index & \makecell{0.29 \\ (<0.001)} & \makecell{-0.11 \\ (0.14)} & \makecell{0.17 \\ (0.02)} & \makecell{-0.04 \\ (0.57)} & \makecell{0.50 \\ (<0.001)} & \makecell{0.15 \\ (0.03)} & -- & \makecell{-0.14 \\ (0.05)} & \makecell{0.47 \\ (<0.001)} & \makecell{0.23 \\ (0.001)} & \makecell{-0.19 \\ (0.008)} & \makecell{0.34 \\ (<0.001)} & \makecell{0.30 \\ (<0.001)} & \makecell{-0.21 \\ (0.004)} & \makecell{-0.17 \\ (0.02)} & \makecell{0.06 \\ (0.39)} & \makecell{0.04 \\ (0.57)} & \makecell{0.09 \\ (0.23)} \\ \addlinespace
dxa\_aog & \makecell{0.26 \\ (<0.001)} & \makecell{0.43 \\ (<0.001)} & \makecell{0.70 \\ (<0.001)} & \makecell{0.35 \\ (<0.001)} & \makecell{0.66 \\ (<0.001)} & \makecell{0.52 \\ (<0.001)} & \makecell{0.47 \\ (<0.001)} & \makecell{0.08 \\ (0.25)} & -- & \makecell{0.43 \\ (<0.001)} & \makecell{0.04 \\ (0.54)} & \makecell{0.88 \\ (<0.001)} & \makecell{0.51 \\ (<0.001)} & \makecell{0.10 \\ (0.15)} & \makecell{0.09 \\ (0.21)} & \makecell{0.45 \\ (<0.001)} & \makecell{0.42 \\ (<0.001)} & \makecell{0.40 \\ (<0.001)} \\ \addlinespace
dxa\_fat & \makecell{0.01 \\ (0.86)} & \makecell{0.72 \\ (<0.001)} & \makecell{0.50 \\ (<0.001)} & \makecell{0.73 \\ (<0.001)} & \makecell{-0.04 \\ (0.56)} & \makecell{0.72 \\ (<0.001)} & \makecell{-0.14 \\ (0.05)} & -- & \makecell{0.08 \\ (0.25)} & \makecell{-0.36 \\ (<0.001)} & \makecell{0.95 \\ (<0.001)} & \makecell{0.17 \\ (0.02)} & \makecell{-0.52 \\ (<0.001)} & \makecell{0.90 \\ (<0.001)} & \makecell{0.95 \\ (<0.001)} & \makecell{0.29 \\ (<0.001)} & \makecell{0.25 \\ (<0.001)} & \makecell{0.10 \\ (0.17)} \\ \addlinespace
dxa\_vos & \makecell{0.53 \\ (<0.001)} & \makecell{0.01 \\ (0.88)} & \makecell{0.24 \\ (0.001)} & \makecell{-0.05 \\ (0.52)} & \makecell{0.35 \\ (<0.001)} & \makecell{0.03 \\ (0.69)} & \makecell{0.23 \\ (0.001)} & \makecell{-0.36 \\ (<0.001)} & \makecell{0.43 \\ (<0.001)} & -- & \makecell{-0.36 \\ (<0.001)} & \makecell{0.38 \\ (<0.001)} & \makecell{0.70 \\ (<0.001)} & \makecell{-0.30 \\ (<0.001)} & \makecell{-0.34 \\ (<0.001)} & \makecell{0.06 \\ (0.42)} & \makecell{0.18 \\ (0.01)} & \makecell{0.35 \\ (<0.001)} \\ \addlinespace
fusion\_pfat & \makecell{0.05 \\ (0.52)} & \makecell{0.78 \\ (<0.001)} & \makecell{0.56 \\ (<0.001)} & \makecell{0.78 \\ (<0.001)} & \makecell{-0.05 \\ (0.49)} & \makecell{0.78 \\ (<0.001)} & \makecell{-0.17 \\ (0.02)} & \makecell{0.95 \\ (<0.001)} & \makecell{0.09 \\ (0.21)} & \makecell{-0.34 \\ (<0.001)} & \makecell{0.99 \\ (<0.001)} & \makecell{0.19 \\ (0.007)} & \makecell{-0.50 \\ (<0.001)} & \makecell{0.96 \\ (<0.001)} & -- & \makecell{0.29 \\ (<0.001)} & \makecell{0.28 \\ (<0.001)} & \makecell{0.12 \\ (0.08)} \\ \addlinespace
hip\_circ\_cm & \makecell{0.13 \\ (0.08)} & \makecell{0.91 \\ (<0.001)} & \makecell{0.83 \\ (<0.001)} & -- & \makecell{0.02 \\ (0.81)} & \makecell{0.88 \\ (<0.001)} & \makecell{-0.04 \\ (0.57)} & \makecell{0.73 \\ (<0.001)} & \makecell{0.35 \\ (<0.001)} & \makecell{-0.05 \\ (0.52)} & \makecell{0.76 \\ (<0.001)} & \makecell{0.43 \\ (<0.001)} & \makecell{-0.13 \\ (0.08)} & \makecell{0.80 \\ (<0.001)} & \makecell{0.78 \\ (<0.001)} & \makecell{0.43 \\ (<0.001)} & \makecell{0.41 \\ (<0.001)} & \makecell{0.31 \\ (<0.001)} \\ \addlinespace
photoscan\_aog & \makecell{0.24 \\ (0.001)} & \makecell{0.54 \\ (<0.001)} & \makecell{0.74 \\ (<0.001)} & \makecell{0.43 \\ (<0.001)} & \makecell{0.64 \\ (<0.001)} & \makecell{0.60 \\ (<0.001)} & \makecell{0.34 \\ (<0.001)} & \makecell{0.17 \\ (0.02)} & \makecell{0.88 \\ (<0.001)} & \makecell{0.38 \\ (<0.001)} & \makecell{0.16 \\ (0.03)} & -- & \makecell{0.49 \\ (<0.001)} & \makecell{0.19 \\ (0.008)} & \makecell{0.19 \\ (0.007)} & \makecell{0.49 \\ (<0.001)} & \makecell{0.46 \\ (<0.001)} & \makecell{0.45 \\ (<0.001)} \\ \addlinespace
photoscan\_fat & \makecell{0.06 \\ (0.41)} & \makecell{0.75 \\ (<0.001)} & \makecell{0.51 \\ (<0.001)} & \makecell{0.76 \\ (<0.001)} & \makecell{-0.09 \\ (0.22)} & \makecell{0.74 \\ (<0.001)} & \makecell{-0.19 \\ (0.008)} & \makecell{0.95 \\ (<0.001)} & \makecell{0.04 \\ (0.54)} & \makecell{-0.36 \\ (<0.001)} & -- & \makecell{0.16 \\ (0.03)} & \makecell{-0.54 \\ (<0.001)} & \makecell{0.93 \\ (<0.001)} & \makecell{0.99 \\ (<0.001)} & \makecell{0.27 \\ (<0.001)} & \makecell{0.27 \\ (<0.001)} & \makecell{0.11 \\ (0.12)} \\ \addlinespace
photoscan\_vos & \makecell{0.23 \\ (0.001)} & \makecell{-0.08 \\ (0.29)} & \makecell{0.20 \\ (0.006)} & \makecell{-0.13 \\ (0.08)} & \makecell{0.39 \\ (<0.001)} & \makecell{-0.05 \\ (0.46)} & \makecell{0.30 \\ (<0.001)} & \makecell{-0.52 \\ (<0.001)} & \makecell{0.51 \\ (<0.001)} & \makecell{0.70 \\ (<0.001)} & \makecell{-0.54 \\ (<0.001)} & \makecell{0.49 \\ (<0.001)} & -- & \makecell{-0.46 \\ (<0.001)} & \makecell{-0.50 \\ (<0.001)} & \makecell{0.10 \\ (0.15)} & \makecell{0.11 \\ (0.13)} & \makecell{0.25 \\ (<0.001)} \\ \addlinespace
waist\_circ\_cm & \makecell{0.25 \\ (<0.001)} & \makecell{0.88 \\ (<0.001)} & -- & \makecell{0.83 \\ (<0.001)} & \makecell{0.49 \\ (<0.001)} & \makecell{0.88 \\ (<0.001)} & \makecell{0.17 \\ (0.02)} & \makecell{0.50 \\ (<0.001)} & \makecell{0.70 \\ (<0.001)} & \makecell{0.24 \\ (0.001)} & \makecell{0.51 \\ (<0.001)} & \makecell{0.74 \\ (<0.001)} & \makecell{0.20 \\ (0.006)} & \makecell{0.61 \\ (<0.001)} & \makecell{0.56 \\ (<0.001)} & \makecell{0.54 \\ (<0.001)} & \makecell{0.51 \\ (<0.001)} & \makecell{0.48 \\ (<0.001)} \\ \addlinespace
waist\_to\_height\_ratio & \makecell{0.21 \\ (0.004)} & \makecell{0.95 \\ (<0.001)} & \makecell{0.88 \\ (<0.001)} & \makecell{0.88 \\ (<0.001)} & \makecell{0.39 \\ (<0.001)} & -- & \makecell{0.15 \\ (0.03)} & \makecell{0.72 \\ (<0.001)} & \makecell{0.52 \\ (<0.001)} & \makecell{0.03 \\ (0.69)} & \makecell{0.74 \\ (<0.001)} & \makecell{0.60 \\ (<0.001)} & \makecell{-0.05 \\ (0.46)} & \makecell{0.82 \\ (<0.001)} & \makecell{0.78 \\ (<0.001)} & \makecell{0.47 \\ (<0.001)} & \makecell{0.45 \\ (<0.001)} & \makecell{0.40 \\ (<0.001)} \\ \addlinespace
waist\_to\_hip\_ratio & \makecell{0.22 \\ (0.002)} & \makecell{0.30 \\ (<0.001)} & \makecell{0.49 \\ (<0.001)} & \makecell{0.02 \\ (0.81)} & -- & \makecell{0.39 \\ (<0.001)} & \makecell{0.50 \\ (<0.001)} & \makecell{-0.04 \\ (0.56)} & \makecell{0.66 \\ (<0.001)} & \makecell{0.35 \\ (<0.001)} & \makecell{-0.09 \\ (0.22)} & \makecell{0.64 \\ (<0.001)} & \makecell{0.39 \\ (<0.001)} & \makecell{-0.01 \\ (0.88)} & \makecell{-0.05 \\ (0.49)} & \makecell{0.28 \\ (<0.001)} & \makecell{0.25 \\ (<0.001)} & \makecell{0.35 \\ (<0.001)} \\
\bottomrule
\end{tabular}
\end{sidewaystable}

\begin{sidewaystable}[p]
\centering
\captionsetup{justification=centering}
\caption{ Feature Correlation of Female Participants}
\label{tab:feature_correlation_female_final}
\tiny
\setlength{\tabcolsep}{1pt}
\begin{tabular}{lcccccccccccccccccc}
\toprule
\textbf{Feature} & \textbf{age} & \makecell{\textbf{body\_mass\_}\\\textbf{index}} & \makecell{\textbf{waist\_}\\\textbf{circ\_cm}} & \makecell{\textbf{hip\_}\\\textbf{circ\_cm}} & \makecell{\textbf{waist\_to\_}\\\textbf{hip\_ratio}} & \makecell{\textbf{waist\_to\_}\\\textbf{height\_ratio}} & \makecell{\textbf{body\_shape\_}\\\textbf{index}} & \makecell{\textbf{dxa\_}\\\textbf{fat}} & \makecell{\textbf{dxa\_}\\\textbf{aog}} & \makecell{\textbf{dxa\_}\\\textbf{vos}} & \makecell{\textbf{photoscan\_}\\\textbf{fat}} & \makecell{\textbf{photoscan\_}\\\textbf{aog}} & \makecell{\textbf{photoscan\_}\\\textbf{vos}} & \makecell{\textbf{bia\_}\\\textbf{pfat}} & \makecell{\textbf{fusion\_}\\\textbf{pfat}} & \textbf{HOMA-IR} & \makecell{\textbf{BP}\\\textbf{Diastolic}} & \makecell{\textbf{BP}\\\textbf{Systolic}} \\
\midrule
BP Diastolic & \makecell{0.16 \\ (0.07)} & \makecell{0.55 \\ (<0.001)} & \makecell{0.57 \\ (<0.001)} & \makecell{0.51 \\ (<0.001)} & \makecell{0.26 \\ (0.003)} & \makecell{0.53 \\ (<0.001)} & \makecell{-0.03 \\ (0.77)} & \makecell{0.44 \\ (<0.001)} & \makecell{0.51 \\ (<0.001)} & \makecell{0.33 \\ (<0.001)} & \makecell{0.49 \\ (<0.001)} & \makecell{0.57 \\ (<0.001)} & \makecell{0.28 \\ (0.001)} & \makecell{0.51 \\ (<0.001)} & \makecell{0.48 \\ (<0.001)} & \makecell{0.39 \\ (<0.001)} & -- & \makecell{0.87 \\ (<0.001)} \\ \addlinespace
BP Systolic & \makecell{0.30 \\ (<0.001)} & \makecell{0.53 \\ (<0.001)} & \makecell{0.54 \\ (<0.001)} & \makecell{0.48 \\ (<0.001)} & \makecell{0.25 \\ (0.004)} & \makecell{0.51 \\ (<0.001)} & \makecell{-0.02 \\ (0.79)} & \makecell{0.38 \\ (<0.001)} & \makecell{0.43 \\ (<0.001)} & \makecell{0.41 \\ (<0.001)} & \makecell{0.43 \\ (<0.001)} & \makecell{0.50 \\ (<0.001)} & \makecell{0.27 \\ (0.002)} & \makecell{0.49 \\ (<0.001)} & \makecell{0.43 \\ (<0.001)} & \makecell{0.32 \\ (<0.001)} & \makecell{0.87 \\ (<0.001)} & -- \\ \addlinespace
HOMA-IR & \makecell{-0.02 \\ (0.82)} & \makecell{0.58 \\ (<0.001)} & \makecell{0.60 \\ (<0.001)} & \makecell{0.56 \\ (<0.001)} & \makecell{0.20 \\ (0.02)} & \makecell{0.55 \\ (<0.001)} & \makecell{0.00 \\ (0.97)} & \makecell{0.45 \\ (<0.001)} & \makecell{0.46 \\ (<0.001)} & \makecell{0.20 \\ (0.02)} & \makecell{0.46 \\ (<0.001)} & \makecell{0.50 \\ (<0.001)} & \makecell{0.22 \\ (0.01)} & \makecell{0.57 \\ (<0.001)} & \makecell{0.48 \\ (<0.001)} & -- & \makecell{0.39 \\ (<0.001)} & \makecell{0.32 \\ (<0.001)} \\ \addlinespace
age & -- & \makecell{0.24 \\ (0.007)} & \makecell{0.31 \\ (<0.001)} & \makecell{0.22 \\ (0.01)} & \makecell{0.21 \\ (0.02)} & \makecell{0.27 \\ (0.002)} & \makecell{0.16 \\ (0.06)} & \makecell{0.15 \\ (0.09)} & \makecell{0.25 \\ (0.003)} & \makecell{0.57 \\ (<0.001)} & \makecell{0.24 \\ (0.005)} & \makecell{0.27 \\ (0.002)} & \makecell{0.08 \\ (0.35)} & \makecell{0.24 \\ (0.007)} & \makecell{0.22 \\ (0.01)} & \makecell{-0.02 \\ (0.82)} & \makecell{0.16 \\ (0.07)} & \makecell{0.30 \\ (<0.001)} \\ \addlinespace
bia\_pfat & \makecell{0.24 \\ (0.007)} & \makecell{0.98 \\ (<0.001)} & \makecell{0.92 \\ (<0.001)} & \makecell{0.92 \\ (<0.001)} & \makecell{0.37 \\ (<0.001)} & \makecell{0.95 \\ (<0.001)} & \makecell{-0.01 \\ (0.89)} & \makecell{0.87 \\ (<0.001)} & \makecell{0.66 \\ (<0.001)} & \makecell{0.28 \\ (0.001)} & \makecell{0.91 \\ (<0.001)} & \makecell{0.77 \\ (<0.001)} & \makecell{0.34 \\ (<0.001)} & -- & \makecell{0.94 \\ (<0.001)} & \makecell{0.57 \\ (<0.001)} & \makecell{0.51 \\ (<0.001)} & \makecell{0.49 \\ (<0.001)} \\ \addlinespace
body\_mass\_index & \makecell{0.24 \\ (0.007)} & -- & \makecell{0.93 \\ (<0.001)} & \makecell{0.93 \\ (<0.001)} & \makecell{0.38 \\ (<0.001)} & \makecell{0.95 \\ (<0.001)} & \makecell{-0.06 \\ (0.49)} & \makecell{0.82 \\ (<0.001)} & \makecell{0.66 \\ (<0.001)} & \makecell{0.31 \\ (<0.001)} & \makecell{0.89 \\ (<0.001)} & \makecell{0.78 \\ (<0.001)} & \makecell{0.34 \\ (<0.001)} & \makecell{0.98 \\ (<0.001)} & \makecell{0.91 \\ (<0.001)} & \makecell{0.58 \\ (<0.001)} & \makecell{0.55 \\ (<0.001)} & \makecell{0.53 \\ (<0.001)} \\ \addlinespace
body\_shape\_index & \makecell{0.16 \\ (0.06)} & \makecell{-0.06 \\ (0.49)} & \makecell{0.11 \\ (0.22)} & \makecell{0.02 \\ (0.82)} & \makecell{0.42 \\ (<0.001)} & \makecell{0.20 \\ (0.02)} & -- & \makecell{-0.02 \\ (0.80)} & \makecell{0.32 \\ (<0.001)} & \makecell{0.04 \\ (0.62)} & \makecell{-0.03 \\ (0.71)} & \makecell{0.14 \\ (0.12)} & \makecell{0.09 \\ (0.28)} & \makecell{-0.01 \\ (0.89)} & \makecell{-0.00 \\ (0.96)} & \makecell{0.00 \\ (0.97)} & \makecell{-0.03 \\ (0.77)} & \makecell{-0.02 \\ (0.79)} \\ \addlinespace
dxa\_aog & \makecell{0.25 \\ (0.003)} & \makecell{0.66 \\ (<0.001)} & \makecell{0.77 \\ (<0.001)} & \makecell{0.59 \\ (<0.001)} & \makecell{0.54 \\ (<0.001)} & \makecell{0.74 \\ (<0.001)} & \makecell{0.32 \\ (<0.001)} & \makecell{0.55 \\ (<0.001)} & -- & \makecell{0.50 \\ (<0.001)} & \makecell{0.60 \\ (<0.001)} & \makecell{0.86 \\ (<0.001)} & \makecell{0.55 \\ (<0.001)} & \makecell{0.66 \\ (<0.001)} & \makecell{0.62 \\ (<0.001)} & \makecell{0.46 \\ (<0.001)} & \makecell{0.51 \\ (<0.001)} & \makecell{0.43 \\ (<0.001)} \\ \addlinespace
dxa\_fat & \makecell{0.15 \\ (0.09)} & \makecell{0.82 \\ (<0.001)} & \makecell{0.75 \\ (<0.001)} & \makecell{0.82 \\ (<0.001)} & \makecell{0.26 \\ (0.003)} & \makecell{0.80 \\ (<0.001)} & \makecell{-0.02 \\ (0.80)} & -- & \makecell{0.55 \\ (<0.001)} & \makecell{0.14 \\ (0.10)} & \makecell{0.92 \\ (<0.001)} & \makecell{0.67 \\ (<0.001)} & \makecell{0.19 \\ (0.03)} & \makecell{0.87 \\ (<0.001)} & \makecell{0.93 \\ (<0.001)} & \makecell{0.45 \\ (<0.001)} & \makecell{0.44 \\ (<0.001)} & \makecell{0.38 \\ (<0.001)} \\ \addlinespace
dxa\_vos & \makecell{0.57 \\ (<0.001)} & \makecell{0.31 \\ (<0.001)} & \makecell{0.37 \\ (<0.001)} & \makecell{0.21 \\ (0.01)} & \makecell{0.29 \\ (<0.001)} & \makecell{0.31 \\ (<0.001)} & \makecell{0.04 \\ (0.62)} & \makecell{0.14 \\ (0.10)} & \makecell{0.50 \\ (<0.001)} & -- & \makecell{0.24 \\ (0.006)} & \makecell{0.45 \\ (<0.001)} & \makecell{0.34 \\ (<0.001)} & \makecell{0.28 \\ (0.001)} & \makecell{0.23 \\ (0.007)} & \makecell{0.20 \\ (0.02)} & \makecell{0.33 \\ (<0.001)} & \makecell{0.41 \\ (<0.001)} \\ \addlinespace
fusion\_pfat & \makecell{0.22 \\ (0.01)} & \makecell{0.91 \\ (<0.001)} & \makecell{0.85 \\ (<0.001)} & \makecell{0.88 \\ (<0.001)} & \makecell{0.30 \\ (<0.001)} & \makecell{0.90 \\ (<0.001)} & \makecell{-0.00 \\ (0.96)} & \makecell{0.93 \\ (<0.001)} & \makecell{0.62 \\ (<0.001)} & \makecell{0.23 \\ (0.007)} & \makecell{0.98 \\ (<0.001)} & \makecell{0.76 \\ (<0.001)} & \makecell{0.28 \\ (0.001)} & \makecell{0.94 \\ (<0.001)} & -- & \makecell{0.48 \\ (<0.001)} & \makecell{0.48 \\ (<0.001)} & \makecell{0.43 \\ (<0.001)} \\ \addlinespace
hip\_circ\_cm & \makecell{0.22 \\ (0.01)} & \makecell{0.93 \\ (<0.001)} & \makecell{0.90 \\ (<0.001)} & -- & \makecell{0.13 \\ (0.15)} & \makecell{0.90 \\ (<0.001)} & \makecell{0.02 \\ (0.82)} & \makecell{0.82 \\ (<0.001)} & \makecell{0.59 \\ (<0.001)} & \makecell{0.21 \\ (0.01)} & \makecell{0.88 \\ (<0.001)} & \makecell{0.70 \\ (<0.001)} & \makecell{0.28 \\ (0.001)} & \makecell{0.92 \\ (<0.001)} & \makecell{0.88 \\ (<0.001)} & \makecell{0.56 \\ (<0.001)} & \makecell{0.51 \\ (<0.001)} & \makecell{0.48 \\ (<0.001)} \\ \addlinespace
photoscan\_aog & \makecell{0.27 \\ (0.002)} & \makecell{0.78 \\ (<0.001)} & \makecell{0.82 \\ (<0.001)} & \makecell{0.70 \\ (<0.001)} & \makecell{0.48 \\ (<0.001)} & \makecell{0.82 \\ (<0.001)} & \makecell{0.14 \\ (0.12)} & \makecell{0.67 \\ (<0.001)} & \makecell{0.86 \\ (<0.001)} & \makecell{0.45 \\ (<0.001)} & \makecell{0.77 \\ (<0.001)} & -- & \makecell{0.50 \\ (<0.001)} & \makecell{0.77 \\ (<0.001)} & \makecell{0.76 \\ (<0.001)} & \makecell{0.50 \\ (<0.001)} & \makecell{0.57 \\ (<0.001)} & \makecell{0.50 \\ (<0.001)} \\ \addlinespace
photoscan\_fat & \makecell{0.24 \\ (0.005)} & \makecell{0.89 \\ (<0.001)} & \makecell{0.83 \\ (<0.001)} & \makecell{0.88 \\ (<0.001)} & \makecell{0.27 \\ (0.002)} & \makecell{0.87 \\ (<0.001)} & \makecell{-0.03 \\ (0.71)} & \makecell{0.92 \\ (<0.001)} & \makecell{0.60 \\ (<0.001)} & \makecell{0.24 \\ (0.006)} & -- & \makecell{0.77 \\ (<0.001)} & \makecell{0.25 \\ (0.005)} & \makecell{0.91 \\ (<0.001)} & \makecell{0.98 \\ (<0.001)} & \makecell{0.46 \\ (<0.001)} & \makecell{0.49 \\ (<0.001)} & \makecell{0.43 \\ (<0.001)} \\ \addlinespace
photoscan\_vos & \makecell{0.08 \\ (0.35)} & \makecell{0.34 \\ (<0.001)} & \makecell{0.40 \\ (<0.001)} & \makecell{0.28 \\ (0.001)} & \makecell{0.24 \\ (0.005)} & \makecell{0.38 \\ (<0.001)} & \makecell{0.09 \\ (0.28)} & \makecell{0.19 \\ (0.03)} & \makecell{0.55 \\ (<0.001)} & \makecell{0.34 \\ (<0.001)} & \makecell{0.25 \\ (0.005)} & \makecell{0.50 \\ (<0.001)} & -- & \makecell{0.34 \\ (<0.001)} & \makecell{0.28 \\ (0.001)} & \makecell{0.22 \\ (0.01)} & \makecell{0.28 \\ (0.001)} & \makecell{0.27 \\ (0.002)} \\ \addlinespace
waist\_circ\_cm & \makecell{0.31 \\ (<0.001)} & \makecell{0.93 \\ (<0.001)} & -- & \makecell{0.90 \\ (<0.001)} & \makecell{0.44 \\ (<0.001)} & \makecell{0.93 \\ (<0.001)} & \makecell{0.11 \\ (0.22)} & \makecell{0.75 \\ (<0.001)} & \makecell{0.77 \\ (<0.001)} & \makecell{0.37 \\ (<0.001)} & \makecell{0.83 \\ (<0.001)} & \makecell{0.82 \\ (<0.001)} & \makecell{0.40 \\ (<0.001)} & \makecell{0.92 \\ (<0.001)} & \makecell{0.85 \\ (<0.001)} & \makecell{0.60 \\ (<0.001)} & \makecell{0.57 \\ (<0.001)} & \makecell{0.54 \\ (<0.001)} \\ \addlinespace
waist\_to\_height\_ratio & \makecell{0.27 \\ (0.002)} & \makecell{0.95 \\ (<0.001)} & \makecell{0.93 \\ (<0.001)} & \makecell{0.90 \\ (<0.001)} & \makecell{0.49 \\ (<0.001)} & -- & \makecell{0.20 \\ (0.02)} & \makecell{0.80 \\ (<0.001)} & \makecell{0.74 \\ (<0.001)} & \makecell{0.31 \\ (<0.001)} & \makecell{0.87 \\ (<0.001)} & \makecell{0.82 \\ (<0.001)} & \makecell{0.38 \\ (<0.001)} & \makecell{0.95 \\ (<0.001)} & \makecell{0.90 \\ (<0.001)} & \makecell{0.55 \\ (<0.001)} & \makecell{0.53 \\ (<0.001)} & \makecell{0.51 \\ (<0.001)} \\ \addlinespace
waist\_to\_hip\_ratio & \makecell{0.21 \\ (0.02)} & \makecell{0.38 \\ (<0.001)} & \makecell{0.44 \\ (<0.001)} & \makecell{0.13 \\ (0.15)} & -- & \makecell{0.49 \\ (<0.001)} & \makecell{0.42 \\ (<0.001)} & \makecell{0.26 \\ (0.003)} & \makecell{0.54 \\ (<0.001)} & \makecell{0.29 \\ (<0.001)} & \makecell{0.27 \\ (0.002)} & \makecell{0.48 \\ (<0.001)} & \makecell{0.24 \\ (0.005)} & \makecell{0.37 \\ (<0.001)} & \makecell{0.30 \\ (<0.001)} & \makecell{0.20 \\ (0.02)} & \makecell{0.26 \\ (0.003)} & \makecell{0.25 \\ (0.004)} \\
\bottomrule
\end{tabular}
\end{sidewaystable}

\begin{sidewaystable}[p]
\centering
\captionsetup{justification=centering}
\caption{Feature Correlation of Male Participants}
\label{tab:feature_correlation_male_final}
\tiny
\setlength{\tabcolsep}{1pt}
\begin{tabular}{lcccccccccccccccccc}
\toprule
\textbf{Feature} & \textbf{age} & \makecell{\textbf{body\_mass\_}\\\textbf{index}} & \makecell{\textbf{waist\_}\\\textbf{circ\_cm}} & \makecell{\textbf{hip\_}\\\textbf{circ\_cm}} & \makecell{\textbf{waist\_to\_}\\\textbf{hip\_ratio}} & \makecell{\textbf{waist\_to\_}\\\textbf{height\_ratio}} & \makecell{\textbf{body\_shape\_}\\\textbf{index}} & \makecell{\textbf{dxa\_}\\\textbf{fat}} & \makecell{\textbf{dxa\_}\\\textbf{aog}} & \makecell{\textbf{dxa\_}\\\textbf{vos}} & \makecell{\textbf{photoscan\_}\\\textbf{fat}} & \makecell{\textbf{photoscan\_}\\\textbf{aog}} & \makecell{\textbf{photoscan\_}\\\textbf{vos}} & \makecell{\textbf{bia\_}\\\textbf{pfat}} & \makecell{\textbf{fusion\_}\\\textbf{pfat}} & \textbf{HOMA-IR} & \makecell{\textbf{BP}\\\textbf{Diastolic}} & \makecell{\textbf{BP}\\\textbf{Systolic}} \\
\midrule
BP Diastolic & \makecell{0.11 \\ (0.37)} & \makecell{0.27 \\ (0.03)} & \makecell{0.33 \\ (0.007)} & \makecell{0.21 \\ (0.10)} & \makecell{0.23 \\ (0.07)} & \makecell{0.32 \\ (0.01)} & \makecell{0.11 \\ (0.37)} & \makecell{0.17 \\ (0.17)} & \makecell{0.33 \\ (0.007)} & \makecell{0.01 \\ (0.94)} & \makecell{0.26 \\ (0.04)} & \makecell{0.35 \\ (0.005)} & \makecell{-0.09 \\ (0.50)} & \makecell{0.30 \\ (0.02)} & \makecell{0.26 \\ (0.04)} & \makecell{0.15 \\ (0.23)} & -- & \makecell{0.74 \\ (<0.001)} \\ \addlinespace
BP Systolic & \makecell{0.25 \\ (0.04)} & \makecell{0.24 \\ (0.06)} & \makecell{0.32 \\ (0.01)} & \makecell{0.06 \\ (0.64)} & \makecell{0.38 \\ (0.002)} & \makecell{0.29 \\ (0.02)} & \makecell{0.12 \\ (0.34)} & \makecell{0.02 \\ (0.86)} & \makecell{0.23 \\ (0.07)} & \makecell{0.20 \\ (0.12)} & \makecell{0.07 \\ (0.58)} & \makecell{0.26 \\ (0.04)} & \makecell{0.07 \\ (0.59)} & \makecell{0.14 \\ (0.28)} & \makecell{0.03 \\ (0.79)} & \makecell{0.16 \\ (0.21)} & \makecell{0.74 \\ (<0.001)} & -- \\ \addlinespace
HOMA-IR & \makecell{0.12 \\ (0.34)} & \makecell{0.35 \\ (0.004)} & \makecell{0.45 \\ (<0.001)} & \makecell{0.23 \\ (0.07)} & \makecell{0.43 \\ (<0.001)} & \makecell{0.39 \\ (0.001)} & \makecell{0.14 \\ (0.26)} & \makecell{0.29 \\ (0.02)} & \makecell{0.55 \\ (<0.001)} & \makecell{-0.09 \\ (0.48)} & \makecell{0.28 \\ (0.03)} & \makecell{0.58 \\ (<0.001)} & \makecell{0.04 \\ (0.73)} & \makecell{0.30 \\ (0.01)} & \makecell{0.27 \\ (0.03)} & -- & \makecell{0.15 \\ (0.23)} & \makecell{0.16 \\ (0.21)} \\ \addlinespace
age & -- & \makecell{-0.07 \\ (0.61)} & \makecell{0.13 \\ (0.31)} & \makecell{-0.06 \\ (0.65)} & \makecell{0.24 \\ (0.06)} & \makecell{0.11 \\ (0.40)} & \makecell{0.51 \\ (<0.001)} & \makecell{-0.14 \\ (0.25)} & \makecell{0.29 \\ (0.02)} & \makecell{0.64 \\ (<0.001)} & \makecell{-0.14 \\ (0.26)} & \makecell{0.21 \\ (0.09)} & \makecell{0.49 \\ (<0.001)} & \makecell{-0.10 \\ (0.42)} & \makecell{-0.18 \\ (0.16)} & \makecell{0.12 \\ (0.34)} & \makecell{0.11 \\ (0.37)} & \makecell{0.25 \\ (0.04)} \\ \addlinespace
bia\_pfat & \makecell{-0.10 \\ (0.42)} & \makecell{0.89 \\ (<0.001)} & \makecell{0.76 \\ (<0.001)} & \makecell{0.75 \\ (<0.001)} & \makecell{0.44 \\ (<0.001)} & \makecell{0.87 \\ (<0.001)} & \makecell{0.03 \\ (0.81)} & \makecell{0.77 \\ (<0.001)} & \makecell{0.50 \\ (<0.001)} & \makecell{-0.25 \\ (0.04)} & \makecell{0.81 \\ (<0.001)} & \makecell{0.57 \\ (<0.001)} & \makecell{-0.32 \\ (0.009)} & -- & \makecell{0.89 \\ (<0.001)} & \makecell{0.30 \\ (0.01)} & \makecell{0.30 \\ (0.02)} & \makecell{0.14 \\ (0.28)} \\ \addlinespace
body\_mass\_index & \makecell{-0.07 \\ (0.61)} & -- & \makecell{0.90 \\ (<0.001)} & \makecell{0.81 \\ (<0.001)} & \makecell{0.57 \\ (<0.001)} & \makecell{0.93 \\ (<0.001)} & \makecell{-0.04 \\ (0.77)} & \makecell{0.64 \\ (<0.001)} & \makecell{0.51 \\ (<0.001)} & \makecell{-0.15 \\ (0.25)} & \makecell{0.74 \\ (<0.001)} & \makecell{0.60 \\ (<0.001)} & \makecell{-0.27 \\ (0.03)} & \makecell{0.89 \\ (<0.001)} & \makecell{0.74 \\ (<0.001)} & \makecell{0.35 \\ (0.004)} & \makecell{0.27 \\ (0.03)} & \makecell{0.24 \\ (0.06)} \\ \addlinespace
body\_shape\_index & \makecell{0.51 \\ (<0.001)} & \makecell{-0.04 \\ (0.77)} & \makecell{0.22 \\ (0.08)} & \makecell{0.08 \\ (0.55)} & \makecell{0.42 \\ (<0.001)} & \makecell{0.28 \\ (0.02)} & -- & \makecell{0.32 \\ (0.009)} & \makecell{0.50 \\ (<0.001)} & \makecell{0.12 \\ (0.35)} & \makecell{0.28 \\ (0.02)} & \makecell{0.38 \\ (0.002)} & \makecell{0.06 \\ (0.63)} & \makecell{0.03 \\ (0.81)} & \makecell{0.23 \\ (0.07)} & \makecell{0.14 \\ (0.26)} & \makecell{0.11 \\ (0.37)} & \makecell{0.12 \\ (0.34)} \\ \addlinespace
dxa\_aog & \makecell{0.29 \\ (0.02)} & \makecell{0.51 \\ (<0.001)} & \makecell{0.70 \\ (<0.001)} & \makecell{0.43 \\ (<0.001)} & \makecell{0.60 \\ (<0.001)} & \makecell{0.65 \\ (<0.001)} & \makecell{0.50 \\ (<0.001)} & \makecell{0.47 \\ (<0.001)} & -- & \makecell{0.01 \\ (0.93)} & \makecell{0.49 \\ (<0.001)} & \makecell{0.81 \\ (<0.001)} & \makecell{0.04 \\ (0.78)} & \makecell{0.50 \\ (<0.001)} & \makecell{0.48 \\ (<0.001)} & \makecell{0.55 \\ (<0.001)} & \makecell{0.33 \\ (0.007)} & \makecell{0.23 \\ (0.07)} \\ \addlinespace
dxa\_fat & \makecell{-0.14 \\ (0.25)} & \makecell{0.64 \\ (<0.001)} & \makecell{0.58 \\ (<0.001)} & \makecell{0.60 \\ (<0.001)} & \makecell{0.44 \\ (<0.001)} & \makecell{0.73 \\ (<0.001)} & \makecell{0.32 \\ (0.009)} & -- & \makecell{0.47 \\ (<0.001)} & \makecell{-0.36 \\ (0.003)} & \makecell{0.91 \\ (<0.001)} & \makecell{0.53 \\ (<0.001)} & \makecell{-0.55 \\ (<0.001)} & \makecell{0.77 \\ (<0.001)} & \makecell{0.91 \\ (<0.001)} & \makecell{0.29 \\ (0.02)} & \makecell{0.17 \\ (0.17)} & \makecell{0.02 \\ (0.86)} \\ \addlinespace
dxa\_vos & \makecell{0.64 \\ (<0.001)} & \makecell{-0.15 \\ (0.25)} & \makecell{-0.00 \\ (0.97)} & \makecell{-0.14 \\ (0.26)} & \makecell{0.02 \\ (0.89)} & \makecell{-0.11 \\ (0.37)} & \makecell{0.12 \\ (0.35)} & \makecell{-0.36 \\ (0.003)} & \makecell{0.01 \\ (0.93)} & -- & \makecell{-0.41 \\ (0.001)} & \makecell{-0.10 \\ (0.45)} & \makecell{0.64 \\ (<0.001)} & \makecell{-0.25 \\ (0.04)} & \makecell{-0.42 \\ (0.001)} & \makecell{-0.09 \\ (0.48)} & \makecell{0.01 \\ (0.94)} & \makecell{0.20 \\ (0.12)} \\ \addlinespace
fusion\_pfat & \makecell{-0.18 \\ (0.16)} & \makecell{0.74 \\ (<0.001)} & \makecell{0.68 \\ (<0.001)} & \makecell{0.71 \\ (<0.001)} & \makecell{0.40 \\ (0.001)} & \makecell{0.78 \\ (<0.001)} & \makecell{0.23 \\ (0.07)} & \makecell{0.91 \\ (<0.001)} & \makecell{0.48 \\ (<0.001)} & \makecell{-0.42 \\ (0.001)} & \makecell{0.98 \\ (<0.001)} & \makecell{0.55 \\ (<0.001)} & \makecell{-0.50 \\ (<0.001)} & \makecell{0.89 \\ (<0.001)} & -- & \makecell{0.27 \\ (0.03)} & \makecell{0.26 \\ (0.04)} & \makecell{0.03 \\ (0.79)} \\ \addlinespace
hip\_circ\_cm & \makecell{-0.06 \\ (0.65)} & \makecell{0.81 \\ (<0.001)} & \makecell{0.80 \\ (<0.001)} & -- & \makecell{0.15 \\ (0.25)} & \makecell{0.74 \\ (<0.001)} & \makecell{0.08 \\ (0.55)} & \makecell{0.60 \\ (<0.001)} & \makecell{0.43 \\ (<0.001)} & \makecell{-0.14 \\ (0.26)} & \makecell{0.72 \\ (<0.001)} & \makecell{0.40 \\ (0.001)} & \makecell{-0.29 \\ (0.02)} & \makecell{0.75 \\ (<0.001)} & \makecell{0.71 \\ (<0.001)} & \makecell{0.23 \\ (0.07)} & \makecell{0.21 \\ (0.10)} & \makecell{0.06 \\ (0.64)} \\ \addlinespace
photoscan\_aog & \makecell{0.21 \\ (0.09)} & \makecell{0.60 \\ (<0.001)} & \makecell{0.69 \\ (<0.001)} & \makecell{0.40 \\ (0.001)} & \makecell{0.65 \\ (<0.001)} & \makecell{0.75 \\ (<0.001)} & \makecell{0.38 \\ (0.002)} & \makecell{0.53 \\ (<0.001)} & \makecell{0.81 \\ (<0.001)} & \makecell{-0.10 \\ (0.45)} & \makecell{0.58 \\ (<0.001)} & -- & \makecell{0.00 \\ (0.97)} & \makecell{0.57 \\ (<0.001)} & \makecell{0.55 \\ (<0.001)} & \makecell{0.58 \\ (<0.001)} & \makecell{0.35 \\ (0.005)} & \makecell{0.26 \\ (0.04)} \\ \addlinespace
photoscan\_fat & \makecell{-0.14 \\ (0.26)} & \makecell{0.74 \\ (<0.001)} & \makecell{0.70 \\ (<0.001)} & \makecell{0.72 \\ (<0.001)} & \makecell{0.45 \\ (<0.001)} & \makecell{0.79 \\ (<0.001)} & \makecell{0.28 \\ (0.02)} & \makecell{0.91 \\ (<0.001)} & \makecell{0.49 \\ (<0.001)} & \makecell{-0.41 \\ (0.001)} & -- & \makecell{0.58 \\ (<0.001)} & \makecell{-0.53 \\ (<0.001)} & \makecell{0.81 \\ (<0.001)} & \makecell{0.98 \\ (<0.001)} & \makecell{0.28 \\ (0.03)} & \makecell{0.26 \\ (0.04)} & \makecell{0.07 \\ (0.58)} \\ \addlinespace
photoscan\_vos & \makecell{0.49 \\ (<0.001)} & \makecell{-0.27 \\ (0.03)} & \makecell{-0.11 \\ (0.38)} & \makecell{-0.29 \\ (0.02)} & \makecell{-0.04 \\ (0.72)} & \makecell{-0.25 \\ (0.04)} & \makecell{0.06 \\ (0.63)} & \makecell{-0.55 \\ (<0.001)} & \makecell{0.04 \\ (0.78)} & \makecell{0.64 \\ (<0.001)} & \makecell{-0.53 \\ (<0.001)} & \makecell{0.00 \\ (0.97)} & -- & \makecell{-0.32 \\ (0.009)} & \makecell{-0.50 \\ (<0.001)} & \makecell{0.04 \\ (0.73)} & \makecell{-0.09 \\ (0.50)} & \makecell{0.07 \\ (0.59)} \\ \addlinespace
waist\_circ\_cm & \makecell{0.13 \\ (0.31)} & \makecell{0.90 \\ (<0.001)} & -- & \makecell{0.80 \\ (<0.001)} & \makecell{0.64 \\ (<0.001)} & \makecell{0.90 \\ (<0.001)} & \makecell{0.22 \\ (0.08)} & \makecell{0.58 \\ (<0.001)} & \makecell{0.70 \\ (<0.001)} & \makecell{-0.00 \\ (0.97)} & \makecell{0.70 \\ (<0.001)} & \makecell{0.69 \\ (<0.001)} & \makecell{-0.11 \\ (0.38)} & \makecell{0.76 \\ (<0.001)} & \makecell{0.68 \\ (<0.001)} & \makecell{0.45 \\ (<0.001)} & \makecell{0.33 \\ (0.007)} & \makecell{0.32 \\ (0.01)} \\ \addlinespace
waist\_to\_height\_ratio & \makecell{0.11 \\ (0.40)} & \makecell{0.93 \\ (<0.001)} & \makecell{0.90 \\ (<0.001)} & \makecell{0.74 \\ (<0.001)} & \makecell{0.70 \\ (<0.001)} & -- & \makecell{0.28 \\ (0.02)} & \makecell{0.73 \\ (<0.001)} & \makecell{0.65 \\ (<0.001)} & \makecell{-0.11 \\ (0.37)} & \makecell{0.79 \\ (<0.001)} & \makecell{0.75 \\ (<0.001)} & \makecell{-0.25 \\ (0.04)} & \makecell{0.87 \\ (<0.001)} & \makecell{0.78 \\ (<0.001)} & \makecell{0.39 \\ (0.001)} & \makecell{0.32 \\ (0.01)} & \makecell{0.29 \\ (0.02)} \\ \addlinespace
waist\_to\_hip\_ratio & \makecell{0.24 \\ (0.06)} & \makecell{0.57 \\ (<0.001)} & \makecell{0.64 \\ (<0.001)} & \makecell{0.15 \\ (0.25)} & -- & \makecell{0.70 \\ (<0.001)} & \makecell{0.42 \\ (<0.001)} & \makecell{0.44 \\ (<0.001)} & \makecell{0.60 \\ (<0.001)} & \makecell{0.02 \\ (0.89)} & \makecell{0.45 \\ (<0.001)} & \makecell{0.65 \\ (<0.001)} & \makecell{-0.04 \\ (0.72)} & \makecell{0.44 \\ (<0.001)} & \makecell{0.40 \\ (0.001)} & \makecell{0.43 \\ (<0.001)} & \makecell{0.23 \\ (0.07)} & \makecell{0.38 \\ (0.002)} \\
\bottomrule
\end{tabular}
\end{sidewaystable}